\documentclass[twocolumn, 5p]{elsarticle}
\usepackage{verbatim}
\usepackage{subfigure}
\usepackage{float}
\usepackage{amsmath}
\usepackage{graphicx}
\usepackage{amssymb}
\usepackage{amsfonts}
\usepackage{subfigure}
\usepackage{algorithm}
\usepackage{algpseudocode}
\usepackage{times}
\usepackage{fullpage}
\usepackage{placeins}
\usepackage[figuresright]{rotating}
\usepackage{epstopdf}
\usepackage{bbm}
\usepackage{xfrac}
\usepackage{color}
\usepackage{nicefrac}
\usepackage{geometry}
\usepackage{wrapfig}
\usepackage{romannum}
\usepackage{url}
\usepackage{lipsum}
\usepackage{flushend}

\graphicspath{{./figures/}}

\usepackage{lineno,hyperref}
\modulolinenumbers[5]

\newcommand{\Integer}{\mathbb{Z}}

\newcommand{\Complex}{\mathbb{C}}

\newcommand{\Expect}{\mathbb{E}}
\newcommand{\Var}{\operatorname{Var}}
\newcommand{\Bias}{\operatorname{Bias}}
\newcommand{\MSE}{\operatorname{MSE}}

\newcommand{\imunit}{\mathrm{j}}
\newcommand{\euler}{\mathrm{e}}

\newcommand{\vg}{\mathbf{g}}
\newcommand{\vgmag}{{r}}
\newcommand{\vgang}{{\alpha}}

\newcommand{\df}{\Delta f}
\newcommand{\dfone}{\Delta f_1}
\newcommand{\dftwo}{\Delta f_2}
\newcommand{\dfang}{\alpha_f}

\newcommand{\ctf}{H_{\phi}}
\newcommand{\phase}{\chi_{\phi}}

\newcommand{\psd}{S}
\newcommand{\hpsd}{\hat{S}}
\newcommand{\hpsdper}{\hpsd^{(\mathrm{p})}}
\newcommand{\hpsdbar}{\hpsd^{(\mathrm{b})}}
\newcommand{\hpsdwel}{\hpsd^{(\mathrm{w})}}
\newcommand{\hpsdmt}{\hpsd^{(\mathrm{mt})}}
\newcommand{\hpsdlp}{\hpsd^{(\mathrm{lp})}}

\newcommand{\hpsdb}{\hpsd_{\e}}
\newcommand{\hpsdblp}{\hpsd_{\e}^{(\mathrm{lp})}}

\newcommand{\y}{\mathbf{y}}
\newcommand{\z}{\mathbf{z}}

\newcommand{\w}{\mathbf{w}}

\newcommand{\e}{\mathbf{e}}
\newcommand{\x}{\mathbf{x}}

\newcommand{\M}{\mathbf{M}}

\newcommand{\rom}[1]{\uppercase\expandafter{\romannumeral #1\relax}}

\newif\ifdraft

\draftfalse
\ifdraft
    \newcommand{\ah}[1]{\textcolor{red}{Ayelet: #1}\xspace}
    \newcommand{\ja}[1]{\textcolor{blue}{Joakim: #1}\xspace}
\else
    \newcommand{\ah}[1]{\textcolor{red}{}}
    \newcommand{\ja}[1]{\textcolor{blue}{}}
\fi

\journal{Ultramicroscopy}

\biboptions{authoryear}

\begin{document}
\pagenumbering{arabic}
\begin{frontmatter}
\title{Reducing Bias and Variance for CTF Estimation in Single Particle Cryo-EM}
\author[add1]{Ayelet Heimowitz\corref{cor1}}
\ead{aheimowitz@math.princeton.edu}
\author[2]{Joakim And\'{e}n}
\ead{janden@flatironinstitute.org}
\author[add1,add3]{Amit Singer}
\ead{amits@math.princeton.edu}
\cortext[cor1]{Corresponding author}
\address[add1]{The Program in Applied and Computational Mathematics, Princeton University, Princeton, NJ}
\address[2]{Center for Computational Biology, Flatiron Institute, New York, NY}
\address[add3]{Department of Mathematics, Princeton University, Princeton, NJ}
%\thanks{This work was partially supported by the Simons Foundation Math+X Investigator Award and the Moore Foundation Data-Driven Discovery Investigator Award.}}

\onecolumn
\begin{abstract}
When using an electron microscope for imaging of particles embedded in vitreous ice, the recorded image, or micrograph, is a significantly degraded version of the tomographic projection of the sample. Apart from noise, the image is affected by the optical configuration of the microscope. This transformation is typically modeled as a convolution with a point spread function. The Fourier transform of this function, known as the contrast transfer function (CTF), is oscillatory, attenuating and amplifying different frequency bands, and sometimes flipping their signs. High-resolution reconstruction requires this CTF to be accounted for, but as its form depends on experimental parameters, it must first be estimated from the micrograph. We present a new method for CTF estimation based on multitaper methods, which reduces bias and variance in the estimate. We also use known properties of the CTF and the background noise power spectrum to further reduce the variance through background subtraction and steerable basis projection. We show that the resulting power spectrum estimates better capture the zero-crossings of the CTF and yield accurate CTF estimates on several experimental micrographs.
\end{abstract}


\end{frontmatter}
%\begin{multicols}{2}
\begingroup
\let\clearpage\relax
\twocolumn

\section{Introduction}
\endgroup

In recent years, single particle cryo-electron microscopy (cryo-EM) has emerged as a leading tool for resolving
the 3D structure of macromolecules from multiple 2D projections of a specimen~\citep{cheng2017cryo}.
In this technique, multiple copies of a particle are embedded in vitreous ice and imaged in an electron microscope.
This yields a set of micrographs, each containing several 2D particle projections.

The micrograph does not contain clean particle projections but is contaminated by several factors, including noise, ice aggregates and carbon film projection.
The noise stems from an inherent limitation on the number of imaging electrons that can be applied to the specimen.
The interference from carbon film and ice aggregates are due to the particular sample preparation techniques used.

The 2D projections in the micrograph are also distorted by convolution with a point spread function.
This point spread function is due to the electron microscope configuration. It attenuates certain frequencies and flips the sign of certain frequency bands.
A 3D density map reconstructed from distorted projections yields an unreliable representation of the particle \citep{frank2996book}.
It is therefore important to estimate the point spread function and account for it during reconstruction.

To estimate these parameters, it is convenient to consider the Fourier transform of the point spread function, known as the \emph{contrast transfer function} (CTF).
This is due to two factors.
First, the CTF has a simple expression in the polar coordinates of the spatial frequency.
Second, its effect is directly visible in the frequency domain where the CTF acts as a pointwise multiplication rather than a convolution \citep{erickson1971ctf}.

CTF estimation is one of the first steps in the single particle cryo-EM pipeline.
Indeed, accounting for the CTF is needed in a variety of tasks, such as particle picking \citep{heimowitz2018apple}, denoising \citep{bhamre2016denoising}, class averaging \citep{scheres2012relion}, ab initio reconstruction \citep{punjani2017cryosparc}, refinement \citep{scheres2012relion, punjani2017cryosparc,tang2007eman2,grant2018cistem} and heterogeneity analysis \citep{scheres2012relion}.

The CTF is typically modeled as a sine function whose argument depends on the spatial frequency and several parameters of the objective lens of the microscope \citep{frank2996book}.
The parameters we focus on in this paper are the defocus and astigmatism of the objective lens as these are unknown and must be estimated from the data. Additionally, the CTF is multiplied by a damping envelope, which suppresses the information in high frequencies~\citep{sorzano2007damping}.

When estimating the CTF parameters, it is common to first estimate the power spectrum of the micrograph.
The observed micrograph image is typically modeled as a CTF-dependent term plus a noise term unaffected by CTF.
The first term corresponds to a noiseless micrograph, that is, the tomographic projection of the sample filtered by the CTF \citet{zhu19977gaussian}.
Modeling the unfiltered and filtered micrographs as 2D random fields, we find that their power spectra are closely related: the latter equals the former multiplied by the squared CTF. This multiplication induces concentric rings, known as Thon rings~\citep{thon1971book}, in the power spectrum of the filtered micrograph (see Fig.~\ref{fig:astigmatism}). 
Estimating the CTF therefore reduces to fitting the parameters of the CTF to the estimated power spectrum. 

The vast majority of CTF estimation methods use a variant of the periodogram when estimating the power spectrum of the micrograph. This is due to its speed and simplicity.
Unfortunately, the periodogram produces a biased and inconsistent estimate of the micrograph power spectrum.

Beyond these issues with the spectral estimators, fitting the CTF model to the estimated power spectrum is complicated by factors such as  the high levels of noise present in the micrograph, {coincidence loss at the detector and more. 
The expected power spectrum equals the power spectrum of the clean, filtered micrograph plus a background term caused by the aforementioned complications. 
The background  masks the true oscillations of the power spectrum of the particle projection. } 
It is therefore important to estimate and remove the background from the estimated power spectrum~\citep{zhu19977gaussian, rohou2015ctffind4, mindell2003ctffind3, kai2016gctf}. 

Assuming that the micrograph power spectrum and the background were both estimated perfectly, the background-subtracted power spectrum equals the power spectrum of the filtered, clean micrograph.
One way to estimate the CTF parameters is then to maximize the correlation of the background-subtracted power spectrum estimate and a squared CTF (or some monotonic function thereof)~\citep{rohou2015ctffind4, mindell2003ctffind3,kai2016gctf, tani1996correlation}.  
Optimizing the correlation then provides an estimate of the defocus and astigmatism.
Another approach identifies a single ring in the estimated power spectrum and uses it to derive a closed-form solution of the CTF parameters~\citet{yan2017single_ring}. In order to formulate this solution, all prior knowledge regarding the spherical aberration must be ignored.

In this paper, we present ASPIRE-CTF, which is a new method for CTF estimation, available as part of the ASPIRE package.\footnote{\url{https://github.com/ComputationalCryoEM/ASPIRE-Python}}
We first estimate the power spectrum using a multitaper estimator \citep{babadi2014multitaper}, further reducing the variance by averaging estimates from multiple regions of the micrograph.
Using this estimated power spectrum, we estimate the background noise spectrum using linear programming (LP).
Instead of using an approximate background model, our scheme ensures that the background-subtracted power spectrum estimate is non-negative and convex.
We also show that the CTF is contained in the span of a small number of steerable basis functions.
Thus, we further reduce the variance in our power spectrum estimate by projecting onto this span.

Given the power spectrum estimate, we provide two solutions for estimating the CTF parameters.
Our first solution is similar to~\citep{rohou2015ctffind4,mindell2003ctffind3, kai2016gctf, tani1996correlation}, where CTF parameters are estimated by maximizing the correlation of the square root of the power spectrum estimate with the absolute value of simulated CTFs.
The second solution uses the spatial frequencies of several zero-crossings.
Since we expect these zero-crossings to coincide with those of the squared CTF, we use them to define an overdetermined system of equations over the CTF parameters that we then solve. We note that, while our first solution is more robust, our second solution is faster to compute.

Our method is experimentally verified in Section~\ref{sec:experiment}. This is done via a
comparison of the defocus estimates with that of~\citet{rohou2015ctffind4, kai2016gctf} on several datasets from the CTF challenge~\citep{marabini2015challenge}.
We show that our power spectrum estimation method is usually in agreement with one of the state-of-the-art methods~\citet{rohou2015ctffind4, kai2016gctf}.

The main contribution of this paper, appearing in Sections \ref{subsubsec:variance1}-\ref{sec:bk}, is our method for estimating the power spectrum of a micrograph. We reduce the variability of the power spectrum estimate, and are therefore  the first to obtain an estimate where several zero-crossing rings of the CTF are easily recovered without additional assumptions.

\begin{figure}[t]
\centering
{\includegraphics[width=0.78\linewidth]{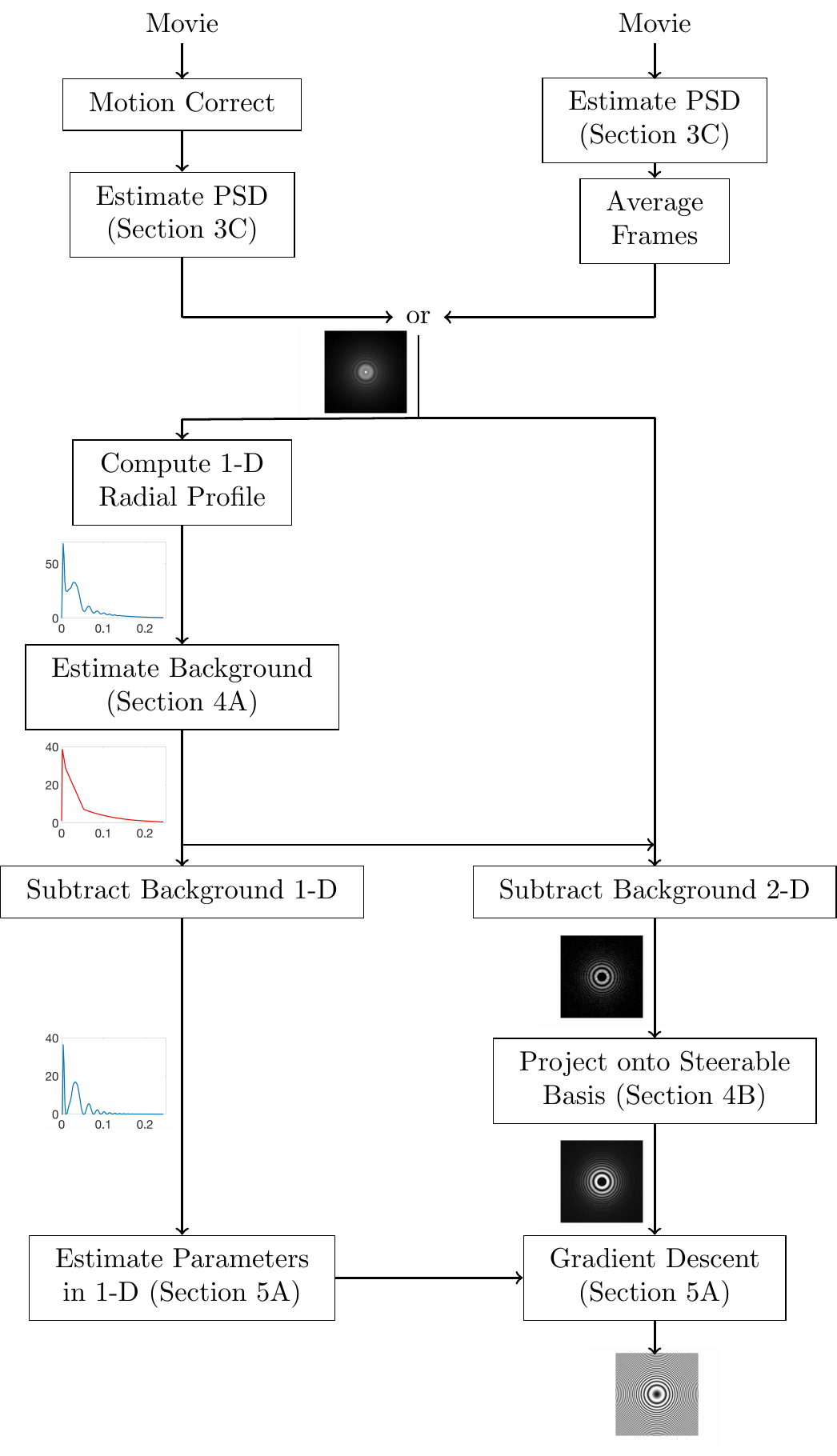}}
\caption{Pipeline of ASPIRE-CTF. The input is a movie and the outputs are the estimated defocus parameters that define the CTF. In the bottom portion of the graph, all actions done on the 1D radial profile of the power spectrum are presented on the left. Additionally, all actions performed on the 2D power spectrum are presented on the right.}
\label{fig:pipeline}
\end{figure}

We present the pipeline of our method in Fig. \ref{fig:pipeline}. For each step of our suggested framework, we refer the reader to the appropriate section of the paper.

\subsection*{Notation}

Given a 2D stationary random field $\x$ defined over $\Integer^2$, we denote its autocovariance function by $R_\x$. 
The Fourier transform of $R_\x$ is known as the \emph{power spectrum} of $\x$ and is given by
\begin{equation}
    \psd_\x(\vg) = \sum_{n_1=-\infty}^\infty \sum_{n_2=-\infty}^\infty R_\x[m, n] \, \euler^{\imunit 2\pi (g_1 n_1 + g_2 n_2)},
\end{equation}
for $\vg = (g_1, g_2) \in [-1/2, 1/2]^2$ and $\imunit = \sqrt{-1}$.
We denote magnitude of the spatial frequency vector $\vg$ by $\vgmag$ and its counterclockwise angle with the positive x-axis by $\vgang$.

\section{Materials and Methods}
\subsection{Problem formulation}
\label{sec:formation}

In the sample preparation stage of the single particle cryo-EM pipeline, many copies of a particle are embedded in vitreous ice.
The imaging process uses an electron microscope to obtain a micrograph containing 2D projections of each instance.
Under the weak-phase object approximation, we may describe this process by the linear model~\citep{frank2996book,thon1971book,mindell2003ctffind3}
\begin{equation}
\label{equ:image_formation}
\y = h_{\phi} * \x + \e,
\end{equation}
where the clean tomographic projection $\x$ and the additive noise $\e$ are modeled as 2D stationary random fields \citep{frank2996book}.
Since convolution preserves stationarity, the observed micrograph $\y$ is also a stationary random field.
In this model, the clean projection $\x$ is convolved with the point spread function of the microscope $h_\phi$ which depends on a parameter vector $\phi$. 
We will at times denote this clean, but filtered, micrograph by $\z = h_\phi * \x$.

\begin{figure}[t]
\centering
\subfigure[]{
{
\setlength{\fboxsep}{0pt}%
\setlength{\fboxrule}{0.3pt}%
\fbox{\includegraphics[width=0.4\linewidth]{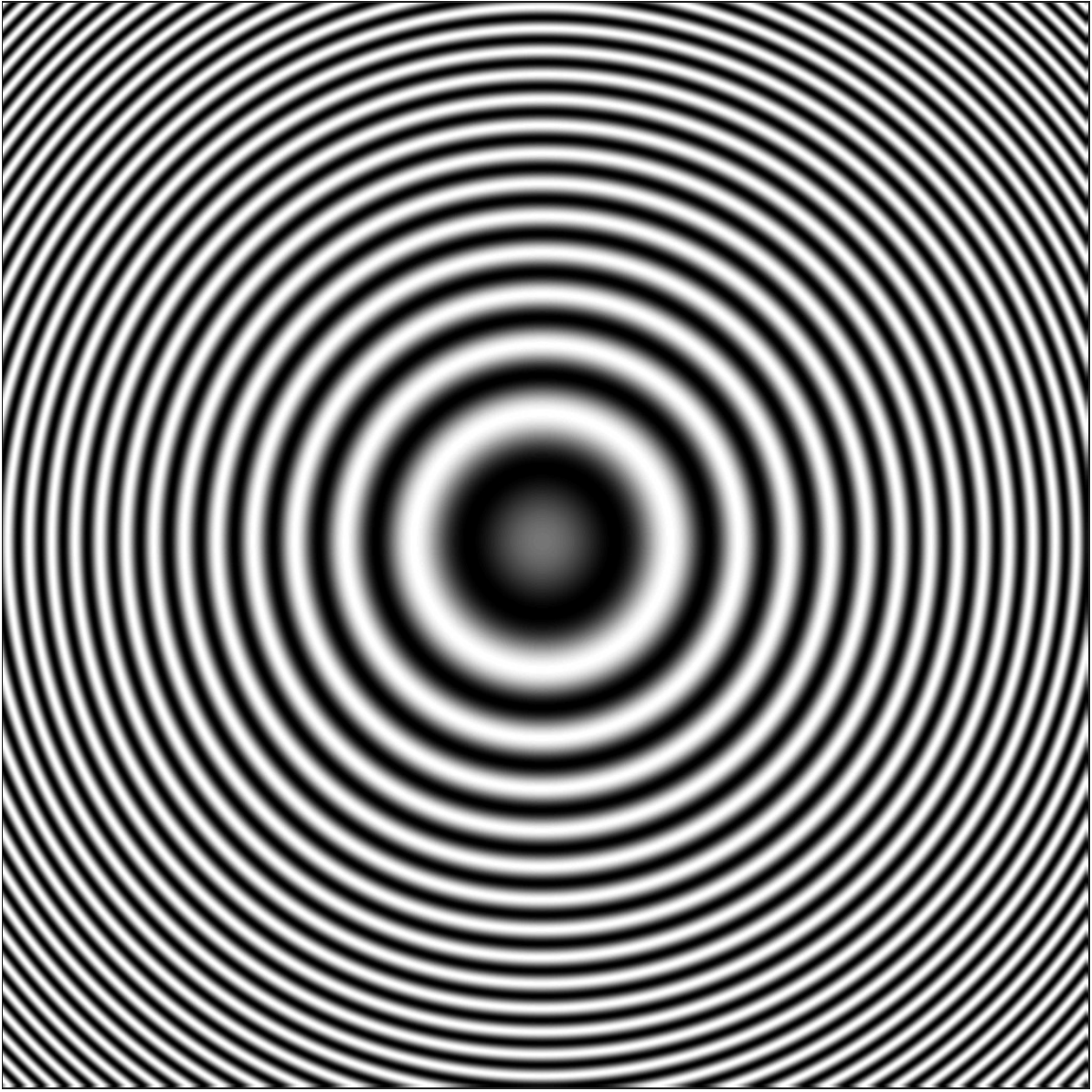}}}
}
\hspace{0.3cm}
\subfigure[]{
{
\setlength{\fboxsep}{0pt}%
\setlength{\fboxrule}{0.3pt}%
\fbox{\includegraphics[width=0.4\linewidth]{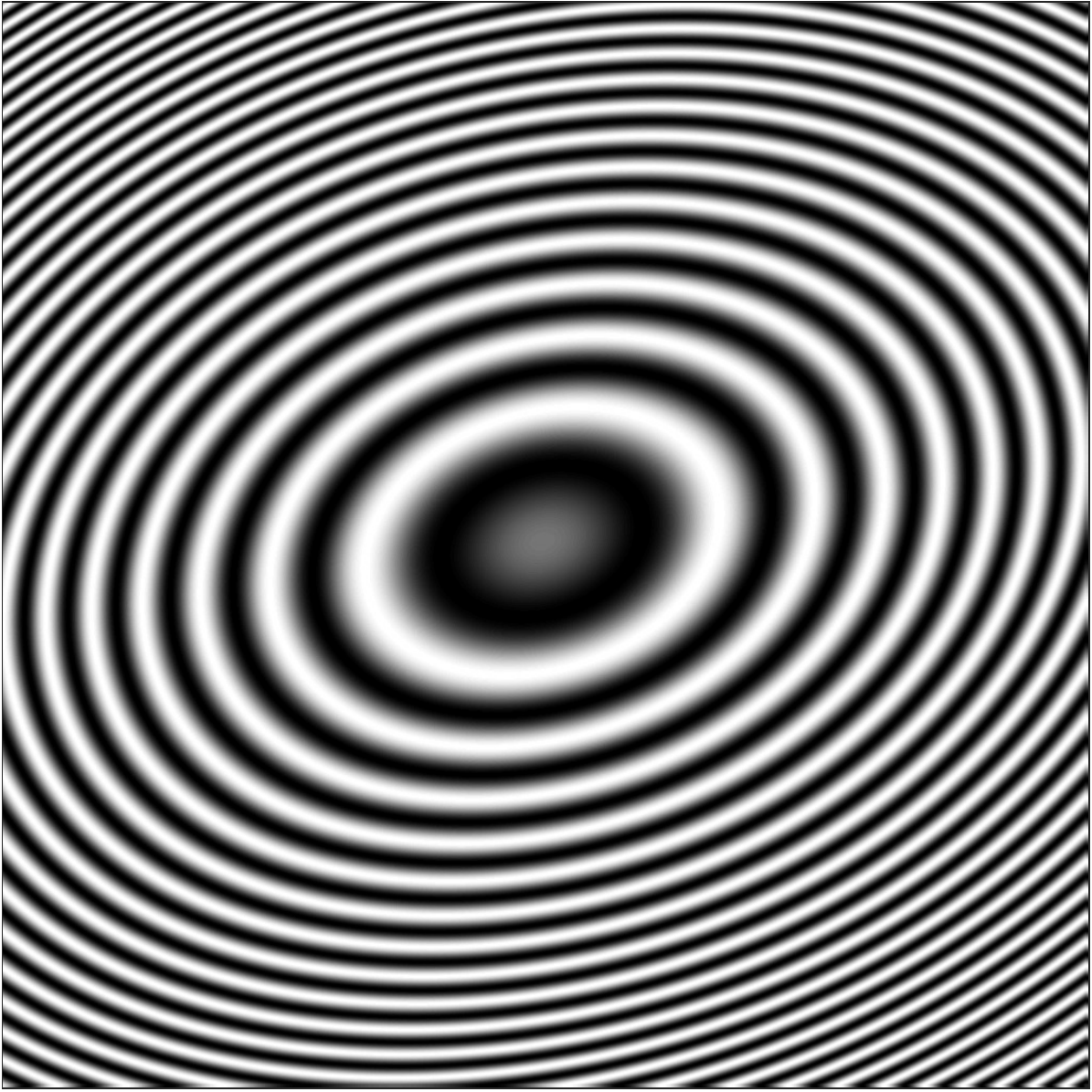}}}
}
\caption{Absolute value of example CTFs. (a) Radially symmetric CTF $\sfrac{(\dfone - \dftwo)}{(\dfone+\dftwo)}=0$. 
(b) Highly astigmatic CTF $\sfrac{(\dfone - \dftwo)}{(\dfone+\dftwo)} = \sfrac{1}{2}$).\vspace{-4ex}}
\label{fig:astigmatism}
\end{figure}

The CTF is the Fourier transform of the point spread function $\ctf$ and may be modeled by~\citep{rohou2015ctffind4}
\begin{equation}
\ctf( \vg ) = - \sin( \phase ( \vg ) ),
\label{equ:ctf_sine}
\end{equation}
where $\vg$ is the spatial frequency.  
Its phase is given by
\begin{equation}
\phase ( \vg ) = \frac{1}{2 p^2} \pi \lambda  \vgmag^2 \df_\phi(\vgang) - \frac{1}{2 p^4} \pi \lambda^3  \vgmag^4 C_s + w,
\end{equation}
where $\lambda$ is the electron wavelength, $C_s$ is the spherical aberration, $w$ is the amplitude contrast, and $p$ is the pixel size.
We also have the astigmatic defocus depth
\begin{equation}
\df_\phi(\vgang) = \dfone + \dftwo + ( \dfone - \dftwo ) \cos ( 2 \vgang - 2 \dfang ),
\label{equ:astigmatism}
\end{equation}
where $\vgang$ is the polar angle of $\vg$ and $\dfone$, $\dftwo$, and $\dfang$ are the major and minor defocus depths and the defocus angle, respectively.
These together form the defocus vector $\phi = (\dfone, \dftwo, \dfang)$, which parametrizes the CTF.
The values $\dfone$ and $\dftwo$ determine the amount of defocus along two perpendicular axes, while $\dfang$ specifies the counterclockwise angle between the major defocus axis and the positive x-axis.
The difference $\dfone - \dftwo$ measures the amount of \emph{astigmatism} in the CTF. A visualization of the effect of astigmatism is provided in Fig. \ref{fig:astigmatism}.

The model~\eqref{equ:ctf_sine} allows us to discern several properties of the CTF.
First, $\ctf$ is real and oscillates between positive and negative values.
As a result, it has several zero crossings.
Second, the CTF is radially symmetric when $\dfone = \dftwo$ (the non-astigmatic case).
Third, with no spherical aberration (i.e., $C_s = 0$) the level sets of the CTF consist of ellipses centered at the origin.
The spherical aberration $C_s$ thus accounts for small deviations from the elliptical shape.

While the parameters $\lambda$, $C_s$, and $w$ are typically known from the microscope configuration, the defocus parameters $\phi$ vary widely between experiments.
We must therefore estimate them to obtain an accurate model of the CTF.

To estimate $\phi$, we turn to the power spectrum of the micrograph.
The power spectra $\psd_\x$, $\psd_\y$, and $\psd_\e$ of $\x$, $\y$, and $\e$, respectively, are related by
\begin{equation}
\label{equ:true_psd}
\psd_\y ( \vg ) = \vert \ctf ( \vg ) \vert^2 \, \psd_\x ( \vg ) + \psd_\e ( \vg ).
\end{equation}
This follows from \eqref{equ:image_formation} and the fact that convolving a stationary random field with $h_\phi$ multiplies its power spectrum by the square Fourier transform magnitude $|\ctf|^2$.

Equation~\eqref{equ:true_psd} suggests that estimates of the power spectra $\psd_\y$, $\psd_\x$, and $\psd_\e$ can be useful in resolving the CTF.
We note that $\psd_\x$ and $\psd_\e$ are slowly decaying while $\vert \ctf \vert^2$ oscillates rapidly in comparison. The background subtracted power spectrum is therefore approximately proportional to $|\ctf|^2$. It follows that in order to 
estimate the defocus parameters $\phi$, we may estimate $\psd_\y-\psd_\e$ and maximize its correlation with $|\ctf|^2$.  
This approach is used in \citep{rohou2015ctffind4, mindell2003ctffind3, kai2016gctf}.

Another approach is to estimate $\phi$ from zero-crossings of $\psd_\y - \psd_\e$ \citep{yan2017single_ring, tani1996correlation}.
Specifically, for spatial frequencies where $\ctf(\vg) = 0$, we have $\psd_\y ( \vg ) - \psd_\e ( \vg ) = 0$.
Identifying these zero-crossings from estimates of $\psd_\y - \psd_\e$ thus constrains the zeros of $\ctf$ and lets us estimate its defocus parameters $\phi$.

For both approaches, the first step is to estimate the background-subtracted power spectrum $\psd_\y - \psd_\e$.
In the following, we propose an estimation method and show how the resulting estimate may be used to estimate $\phi$ by either maximizing correlation or matching zero-crossings.

As mentioned above, the CTF is also multiplied by an exponentially decreasing envelope function \citep{sorzano2007damping}, which effectively acts as a low-pass filter on $h_{\phi} * \x$. 
In this paper, rather than include the envelope function in our analysis, we ignore high frequencies as they are strongly attenuated by the envelope. We also reduce the effect of the envelope function by estimating the CTF using the square root of our power spectrum estimate as in \citep{rohou2015ctffind4, mindell2003ctffind3}.
 In this way, the effect of the envelope function on the two methods discussed  is smaller.
\subsection{Power spectrum estimation}
\label{subsec:estimate}

In this section we present several methods for estimating the power spectrum of the micrograph. We first present the periodogram estimator and then show different methods for reducing its bias and variance.

\subsubsection{Periodogram estimator}
\label{subsec:periodogram}

In an experimental setting, we only have access to an $N \times N$ sample of $\y$, given by the values $\y[k_1, k_2]$ for $(k_1, k_2) \in \{0, 1, \ldots, N-1\}^2$.
Given these values, a common power spectrum estimator is provided by the periodogram \citep{oppenheim1989discrete}
\begin{equation} \label{equ:periodogram}
\hpsdper_{\y} (\vg) = \frac{1}{N^2}  \left\vert \sum_{k_1,k_2=0}^{N-1} \y \left[ k_1, k_2 \right] \, \euler^{-\imunit 2 \pi ( g_1 k_1 + g_2 k_2 ) }  \right\vert^2,
\end{equation}
for $\vg \in [-1/2, 1/2]^2$.
While $\hpsdper_\y(\vg)$ may be calculated for any $\vg$, it is typically calculated on the $N \times N$ grid
\begin{equation}
M_N = \left\{ -\frac{1}{2}, -\frac{1}{2} + \frac{2}{N}, \ldots, \frac{1}{2} - \frac{2}{N} \right\}^2.
\end{equation}
This enables the use of fast Fourier transforms (FFTs) for computing the periodogram with $O(N^2 \log N)$ computational complexity.
Due to this and its ease of implementation, the periodogram is a popular spectral estimator in cryo-EM.

Since our goal is to estimate $\psd_\y$, let us consider how well it is estimated by the periodogram.
The mean square error (MSE) of $\hpsdper_\y$ at $\vg$ is given by
\begin{equation}
\MSE(\hpsdper_{\y}(\vg)) = \Expect\left[\vert \hpsdper_\y(\vg) - \psd_\y(\vg) \vert^2\right].
\end{equation}
To analyze the source of error, it is useful to define the \emph{bias} and \emph{variance} of the periodogram.
The bias is defined as
\begin{equation}
\Bias(\hpsdper_{\y}(\vg)) = \Expect\left[\hpsdper_{\y}(\vg)\right] - \psd_{\y}(\vg)
\label{equ:bias}
\end{equation}
and measures the deviation of the expectation from the true value, while the variance
\begin{equation}
\Var(\hpsdper_{\y}(\vg)) = \Expect\left[\left\vert \hpsdper_\y(\vg) - \Expect\left[\hpsdper_\y(\vg)\right] \right\vert^2 \right]
\end{equation}
measures the average deviation of the periodogram from its expectation.
Both contribute to the MSE through the identity
\begin{equation}
\MSE(\hpsdper_\y(\vg)) = \Bias^2(\hpsdper_\y(\vg)) + \Var(\hpsdper_\y(\vg)).
\end{equation}
A low MSE therefore requires low bias and low variance.

The periodogram, however, fails on both counts.
First, while the periodogram is asymptotically unbiased~\citep{percival1993multitapers,thompson1982spectral}, its bias remains large for small samples.
Second, {the periodogram is an \emph{inconsistent} estimator, that is, its variance does not decrease with an increase in sample size. Therefore, 
a periodogram that extends over the entire micrograph will have variance approximately equal to that of a periodogram that extends over some section of the micrograph.} 
 
In the following sections, we will therefore consider different approaches to reducing both the bias and variance of the periodogram.

\subsubsection{Bartlett's method}
\label{subsubsec:variance1}

We first consider an approach for reducing variance called \emph{Bartlett's method} \citep{oppenheim1989discrete}.
In this approach, the periodogram estimate is computed for several non-overlapping regions of the image.
These estimates are then averaged, reducing the variance by a factor approximately equal to the number of regions used.
It may therefore be tempting to drastically reduce the size of these regions.
However, in experimental data, averaging over regions that are too small will increase the bias.
Among other things, this would prevent us from properly estimating the low spatial frequencies.

We thus divide our image into $B$ non-overlapping blocks $\y_0, \ldots, \y_{B-1}$ of size $K \times K$.
The averaged periodogram is
\begin{equation} \label{equ:periodogram_blocks}
\hpsdbar_\y(\vg) = \frac{1}{B} \sum_{b=0}^{B-1} \hpsdper_{\y_b} (\vg).
\end{equation}
If each block $\y_b$ is independent of the others, we have $\Var(\hpsdbar_\y(\vg)) = B^{-1} \Var(\hpsdper_\y(\vg))$.
Note that, since the block size is now $K \times K$, we sample $\vg$ on $M_K$.

\subsubsection{Welch's method}
\label{subsec:bias}

The expected value of the periodogram estimator is known to be a convolution between the true power spectrum of the micrograph and a 2D Fej\'{e}r kernel~\citep{percival1993multitapers}. As the Fej\'{e}r kernel has high sidelobes, this convolution leads to frequency leakage and therefore a high bias. 

One method of lowering the bias of the periodogram estimation is tapering \citep{percival1993multitapers}. 
This multiplies the data $\y$ by a data taper $\w$ prior to computing the periodogram, resulting in a \emph{modified periodogram}. 
While many options for data tapers exist, such as the Hann window \citep{vulovic2012taper}, Babadi and Brown \citep{babadi2014multitaper} suggest the use of the zeroth-order discrete prolate spheroidal sequence (DPSS) \citep{slepian1978prolates}. 
The expected value of this modified periodogram is  a convolution between the true power spectrum of the micrograph and a kernel with smaller sidelobes than those of the Fej\'{e}r kernel~\citep{percival1993multitapers}. This reduces the frequency leakage, and, therefore, the bias of the estimator.
 
While the taper may be applied to the entire micrograph, it is also possible to apply it to each block in Bartlett's method \eqref{equ:periodogram_blocks}.
The resulting approach is known as \emph{Welch's method} \citep{welch1967wosa}.
Welch also showed that further variance reduction is possible using overlapping (typically half-overlapping) blocks~\citep{frenandez19977periodogram, huang2003env, frank2996book, zhu19977gaussian}.
This yields the \emph{modified averaged periodogram},
\begin{equation}
\hpsdwel_{\y}(\vg) = \frac{1}{B} \sum_{b=1}^{B} \hpsdper_{\y_b \cdot \w}(\vg)
\end{equation}
where $\y_b \cdot \w$ is the pointwise multiplication of $\y_b$ and $\w$.

\subsubsection{Multitaper estimators}
\label{subsec:multi}
As discussed in Section~\ref{subsubsec:variance1}, one way to lower the variance in the periodogram is to average several estimates.
For this reason, Thomson \citep{thompson1982spectral} suggested combining the estimates obtained from multiple tapers.
Each taper yields a different estimate of the power spectrum, and averaging them significantly reduces the variance.
A large number of tapers, however, results in significant smoothing of the power spectrum estimate, so the variance reduction needs to be balanced with an increase in bias for non-smooth power spectra.
Thomson found that higher-order DPSSs were well-suited to this task and called the resulting power spectrum estimator the \emph{multitaper estimator}.
These estimators have recently demonstrated their usefulness for noise power spectrum estimation in cryo-EM \citep{anden2017factor,anden2019multitaper}.  For details of the DPSS data tapers we refer the reader to Appendix A.

Combining all the above methods for variance and bias reduction, we arrive at the multitaper estimator
\begin{equation} \label{equ:multitaper}
\hpsdmt_{\y}(\vg) = \frac{1}{L B} \sum_{b=0}^{B-1} \sum_{\ell=0}^{L-1} \hpsdper_{{\y}_b\cdot\w_\ell}(\vg),
\end{equation}
where $\w_\ell$ is the $\ell$th out $L$ DPSSs for grids of size $K \times K$.

\begin{figure*}
\begin{tabular}{cccc}
\centering
Periodogram & Bartlett's method & Welch's method & Multitaper method\\
{\includegraphics[width=0.22\linewidth]{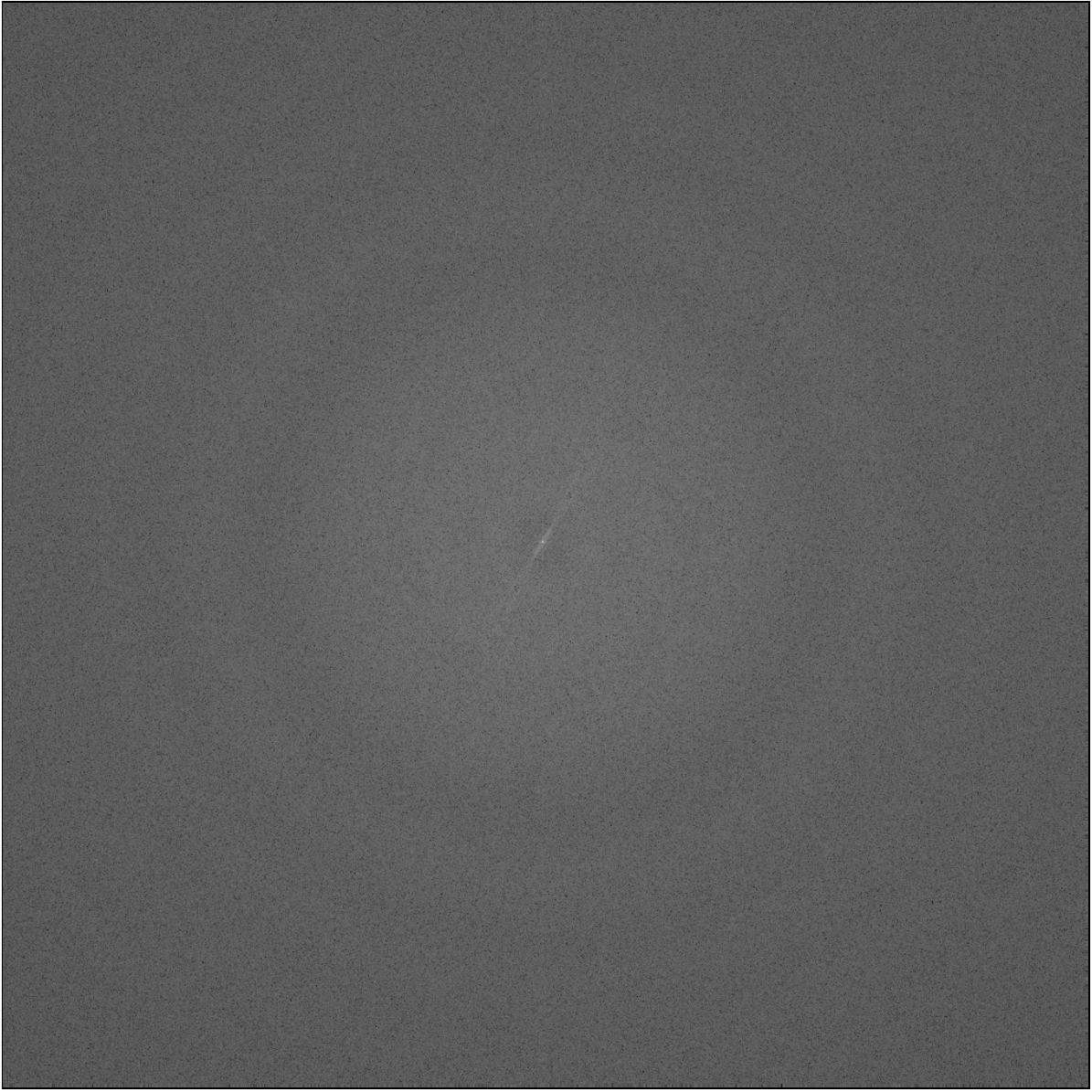}} &
{\includegraphics[width=0.22\linewidth]{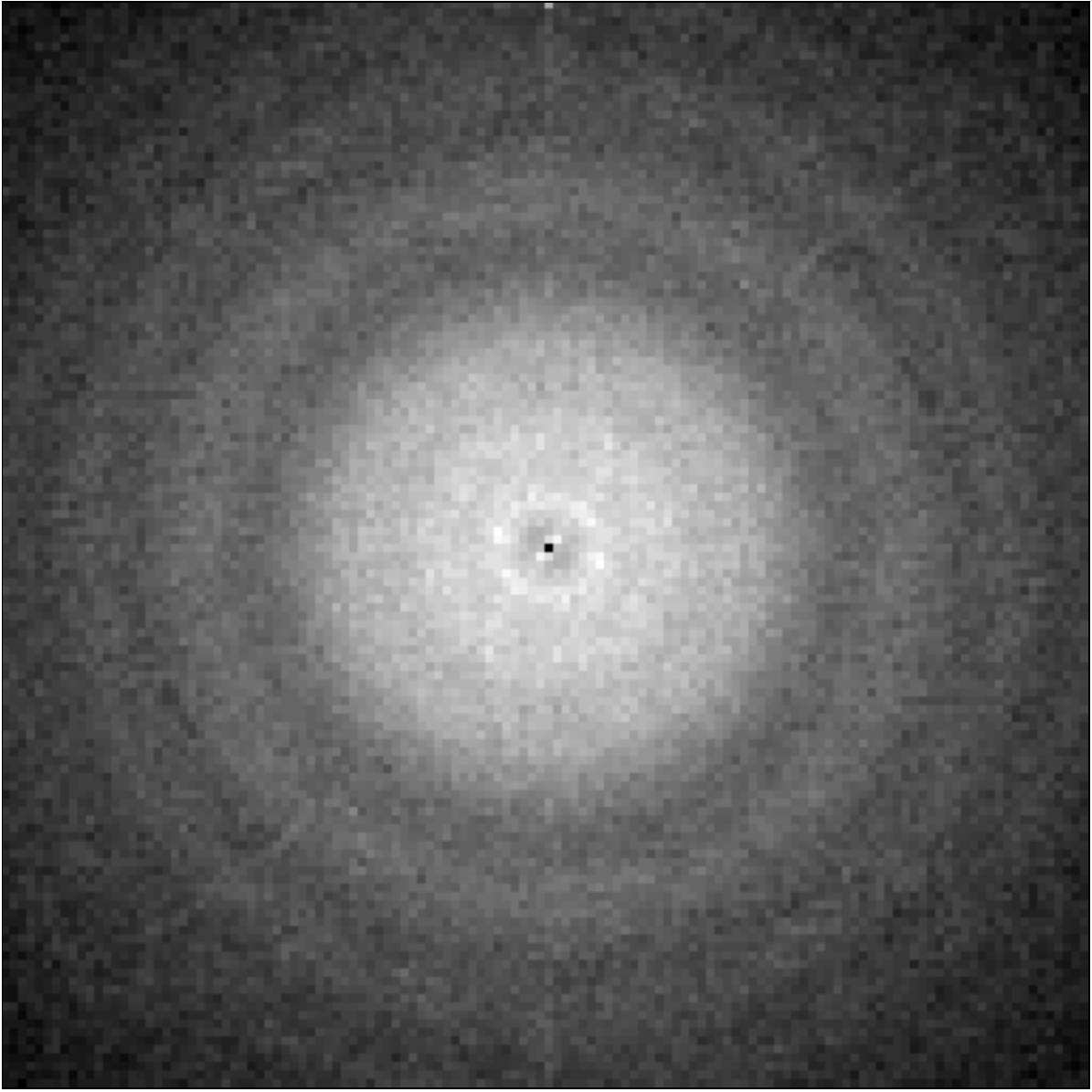}} &
{\includegraphics[width=0.22\linewidth]{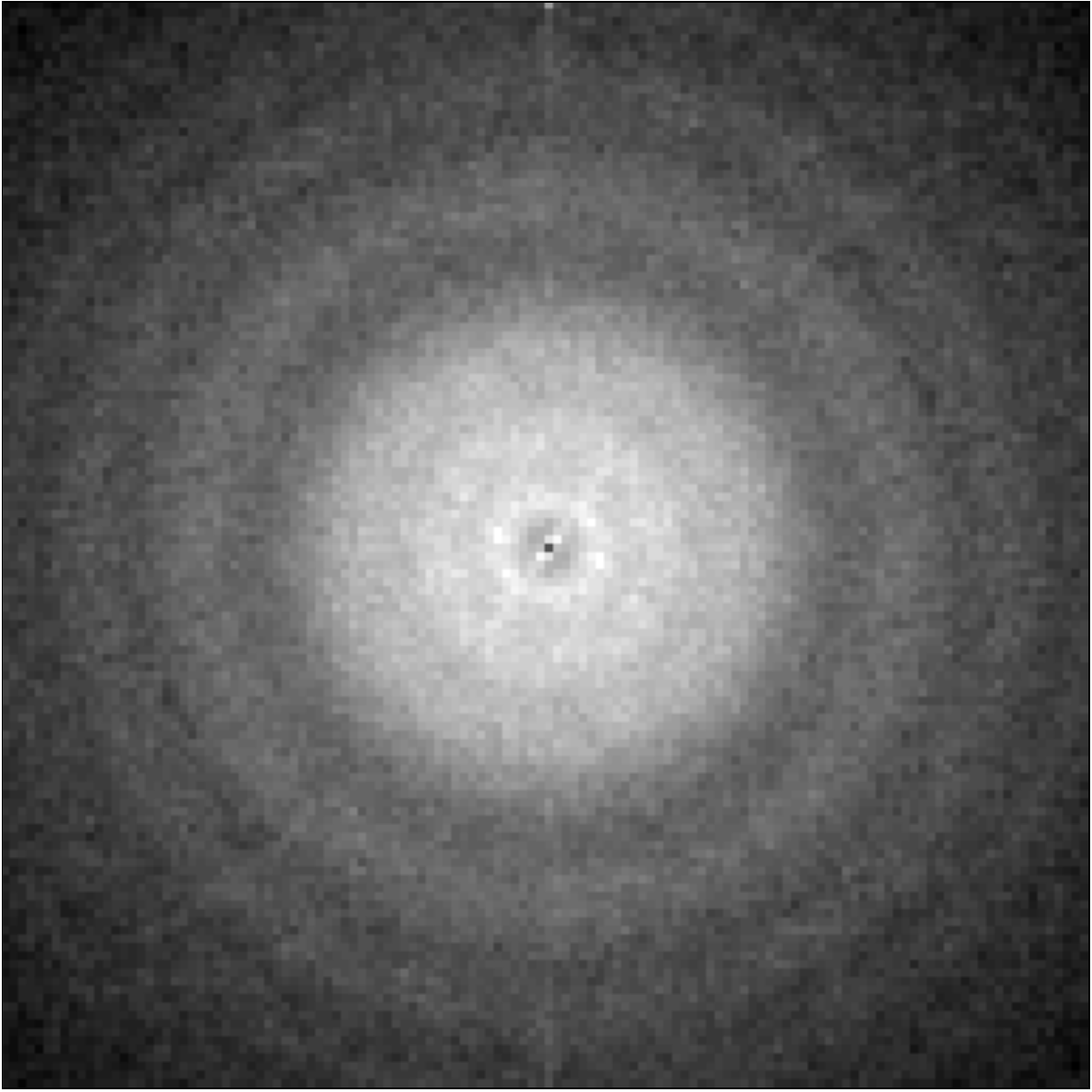}} &
{\includegraphics[width=0.22\linewidth]{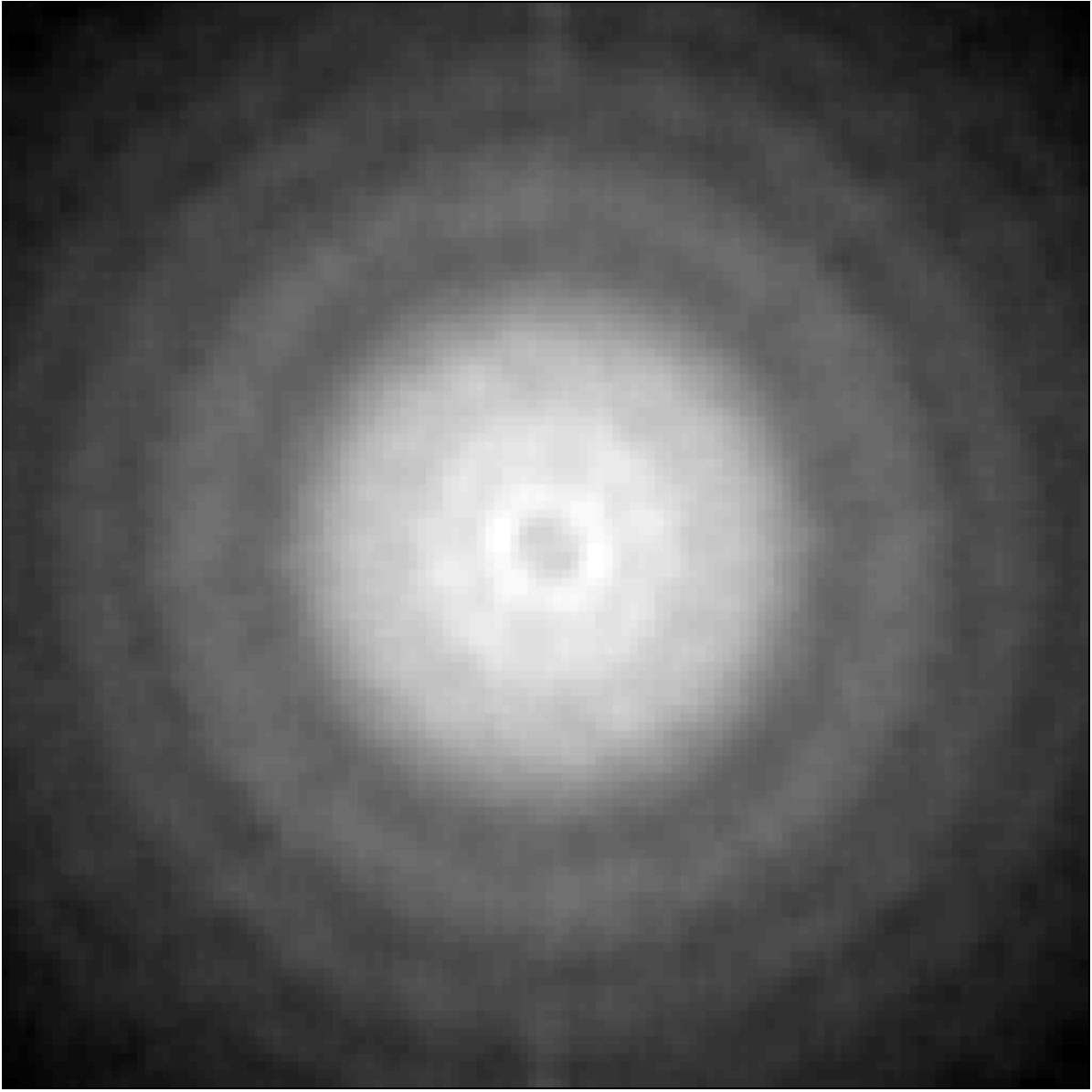}}\\
{\includegraphics[width=0.22\linewidth]{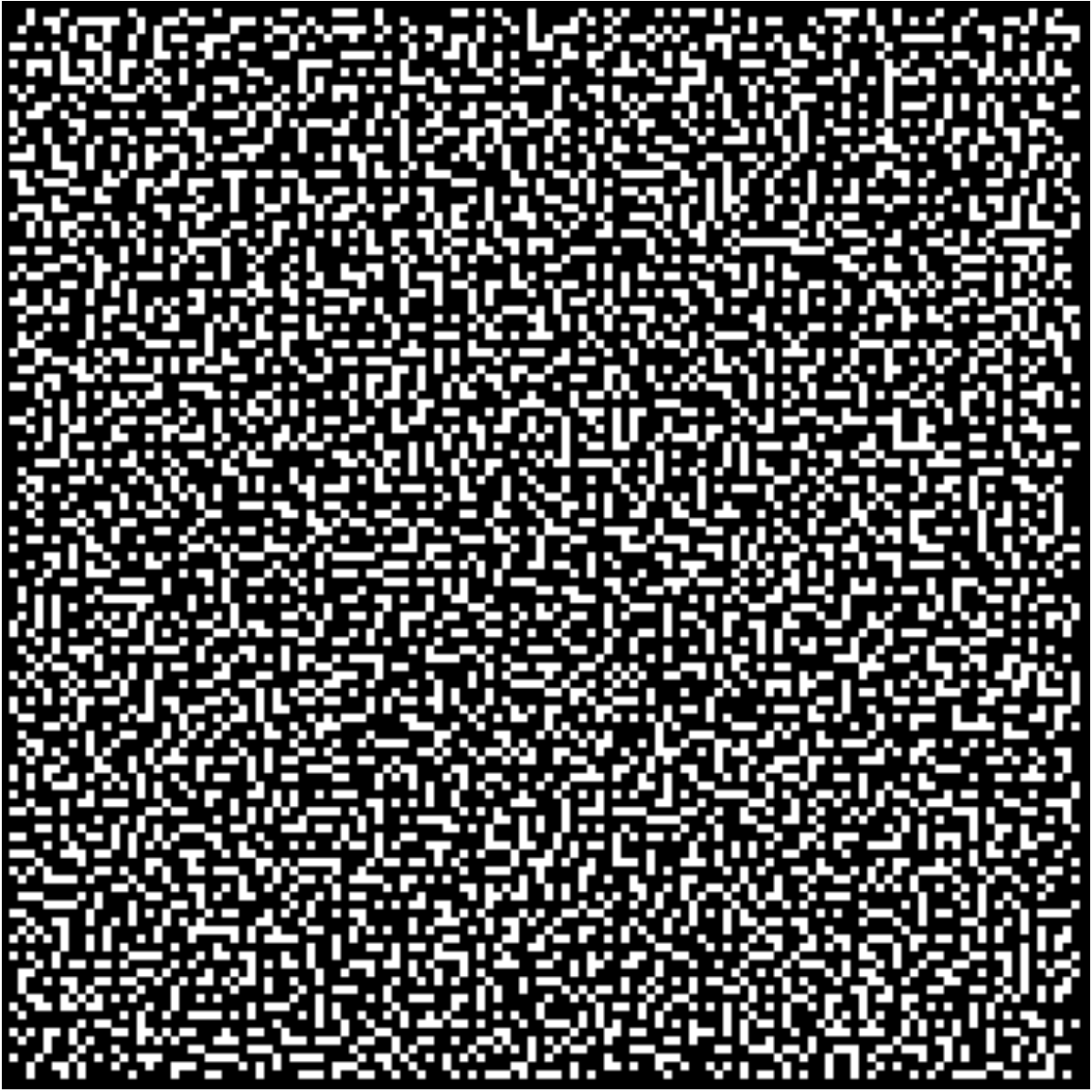}} &
{\includegraphics[width=0.22\linewidth]{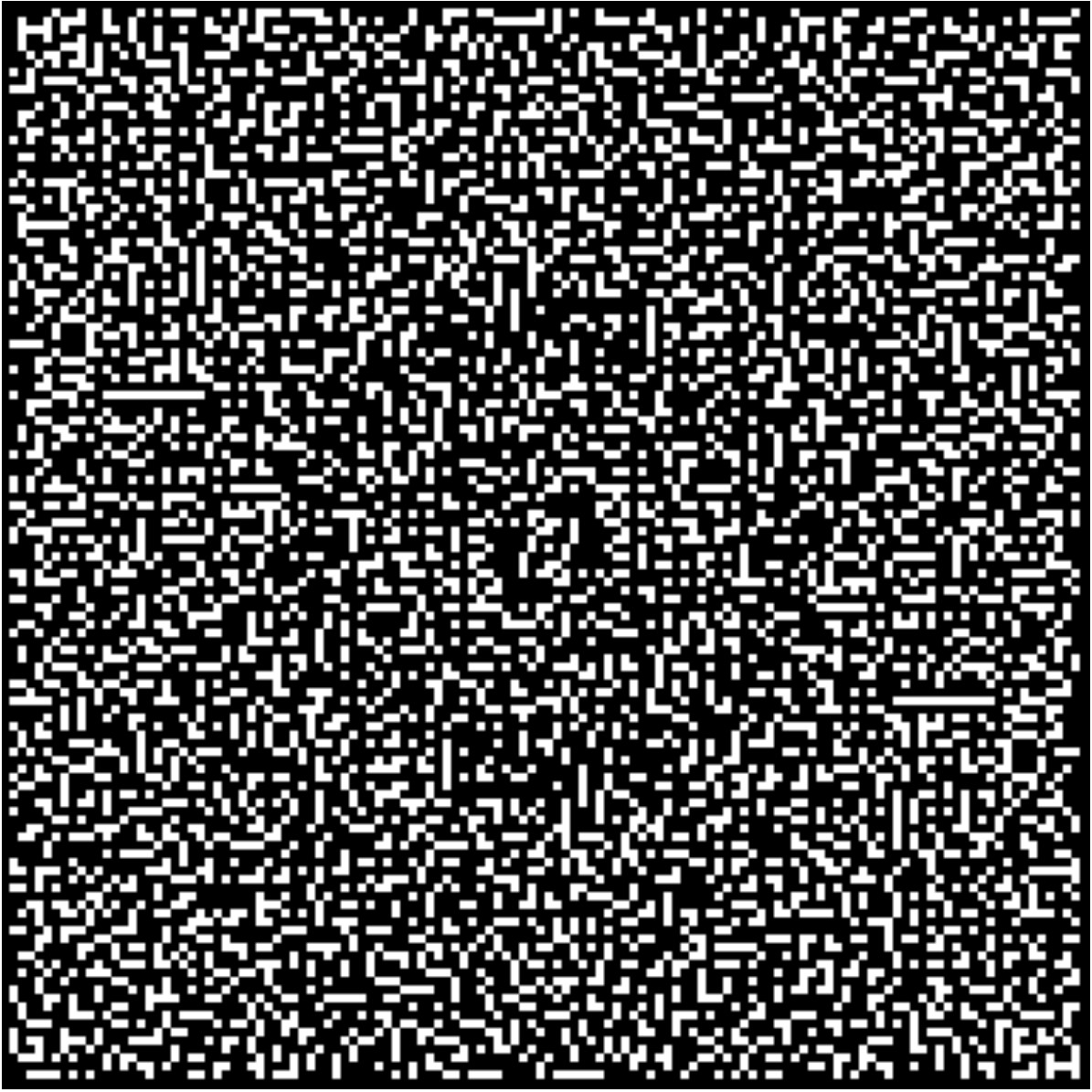}} &
{\includegraphics[width=0.22\linewidth]{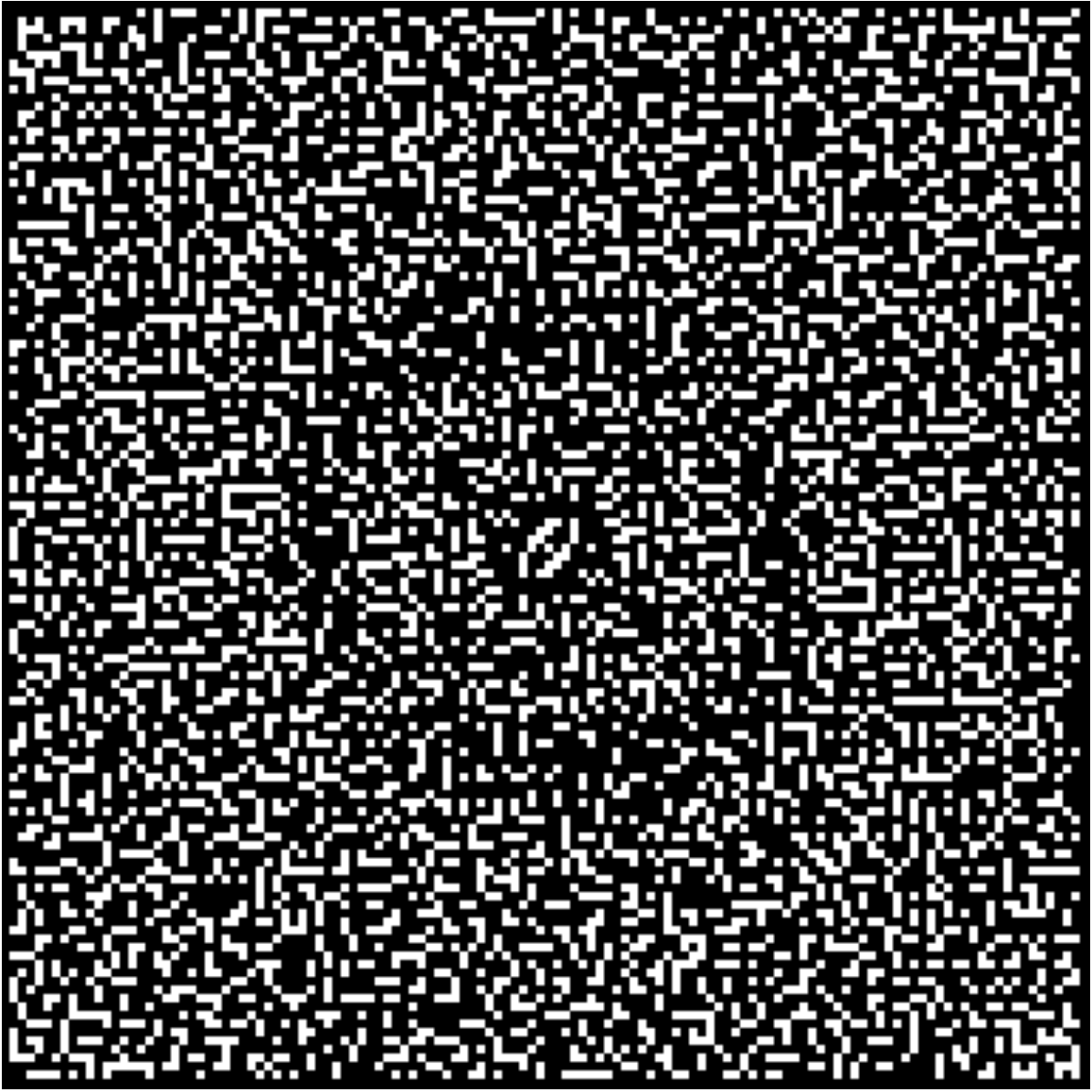}} &
{\includegraphics[width=0.22\linewidth]{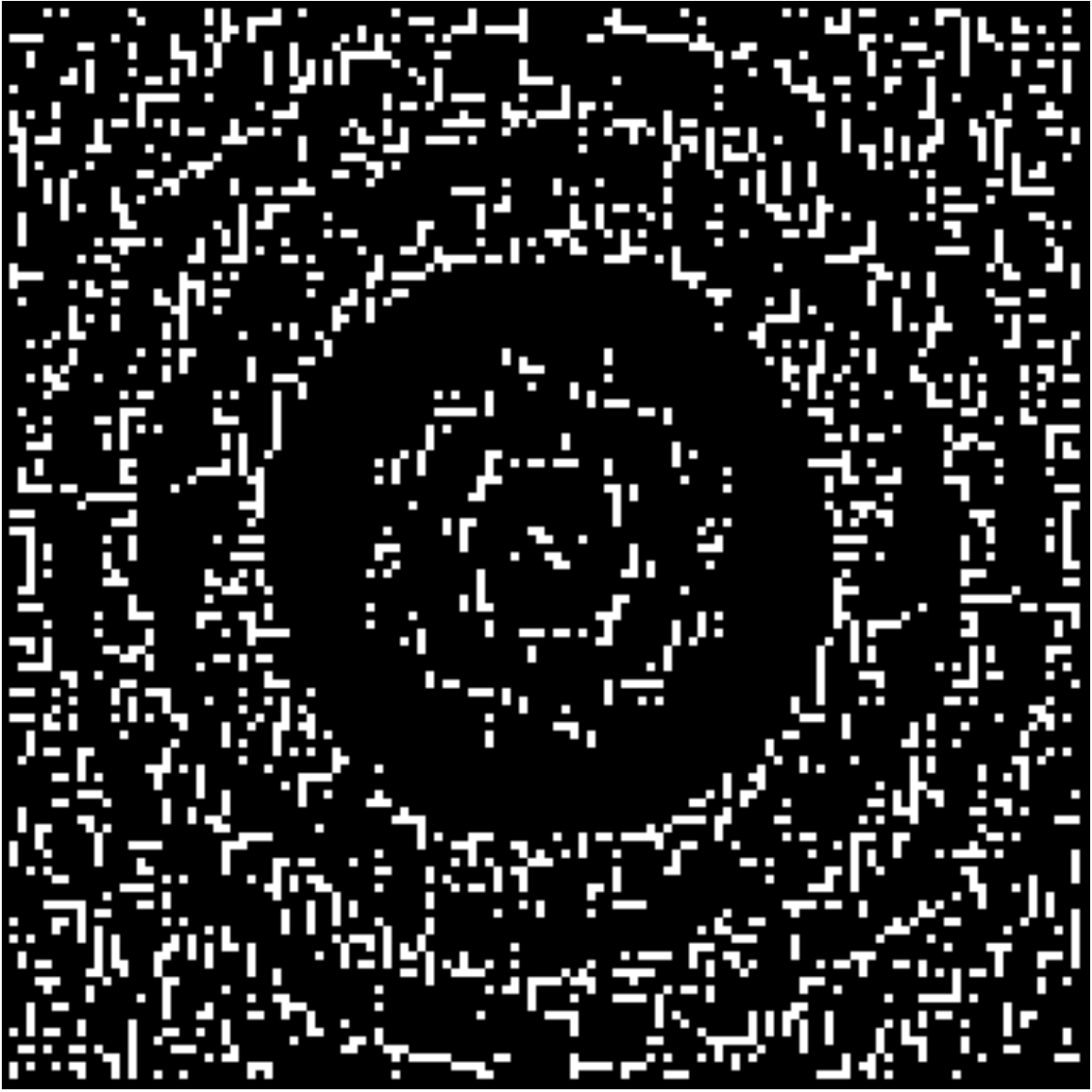}}
\end{tabular}
\caption{   
Power spectrum estimation and zero-crossings of the estimation for a $\beta$-galactosidase micrograph from the EMPIAR-10017 dataset \citep{scheres2015semi}. In the top row we present several power spectrum estimates (plotted on a logarithmic scale). The estimates are produced using the Periodogram estimator (leftmost column), Bartlett's method (center-left column), Welch's method (center-right column) and the Multitaper method with $L=9$ data tapers (rightmost column). Below each estimate we present its zero-crossings (which are determined as specified in Section~\ref{subsec:zero_cross}). 
Blocks of size $512 \times 512$ were used.
The zero-crossings of the CTF are most easily identifiable in the multitaper estimates.}
\label{fig:compare}
\end{figure*}

\begin{figure*}
\begin{tabular}{cccc}
\centering
Periodogram & Bartlett's method & Welch's method & Multitaper method\\
{\includegraphics[width=0.22\linewidth]{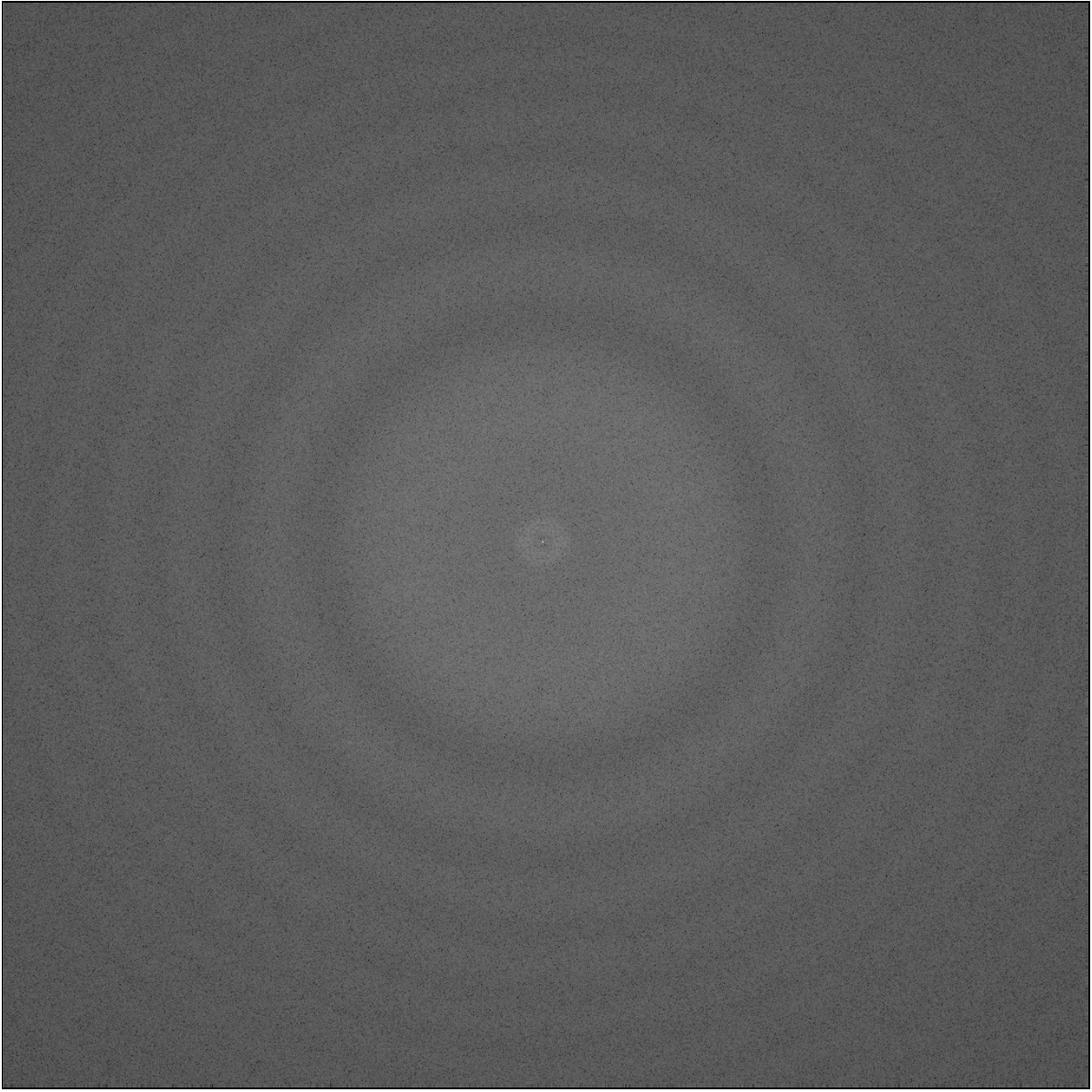}} &
{\includegraphics[width=0.22\linewidth]{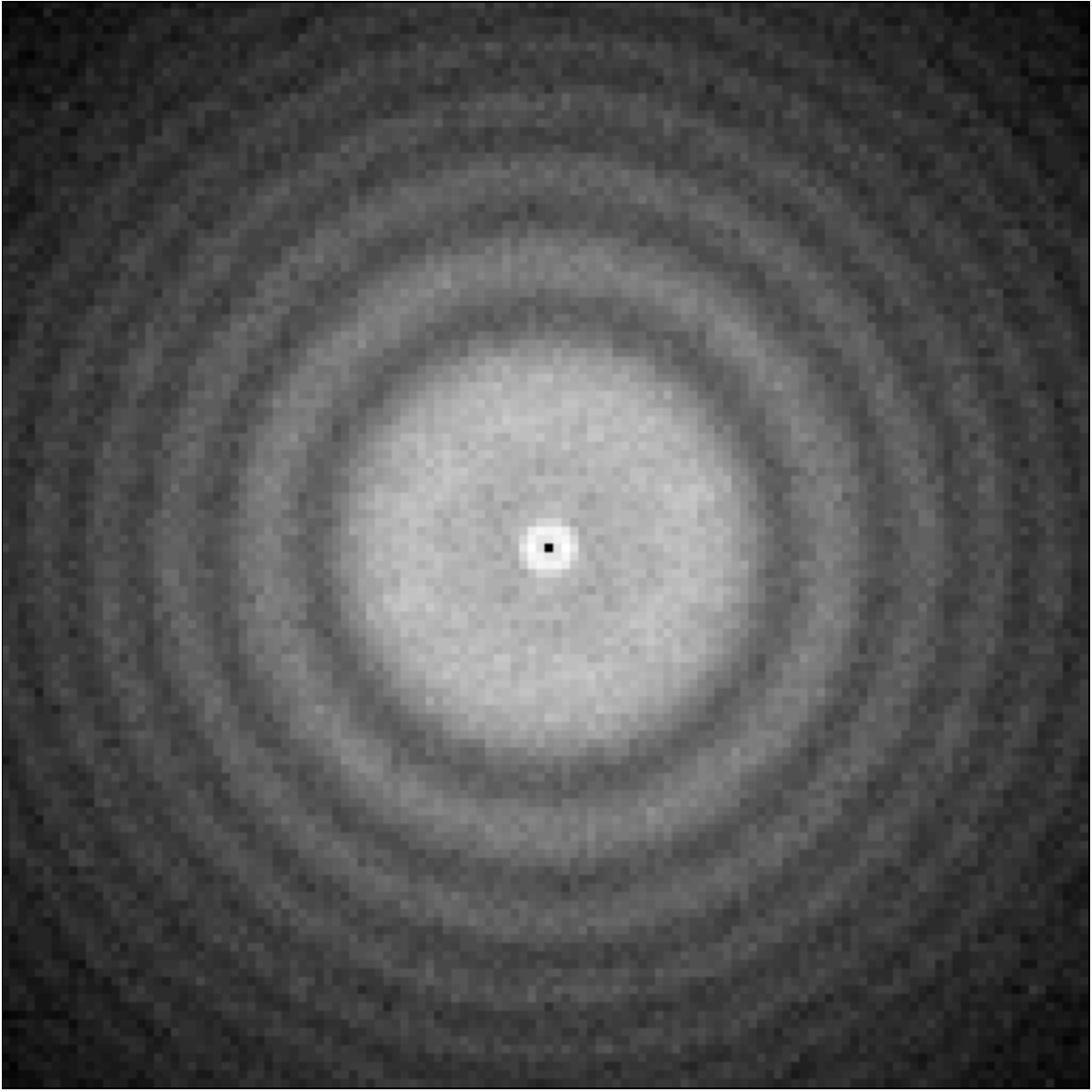}} &
{\includegraphics[width=0.22\linewidth]{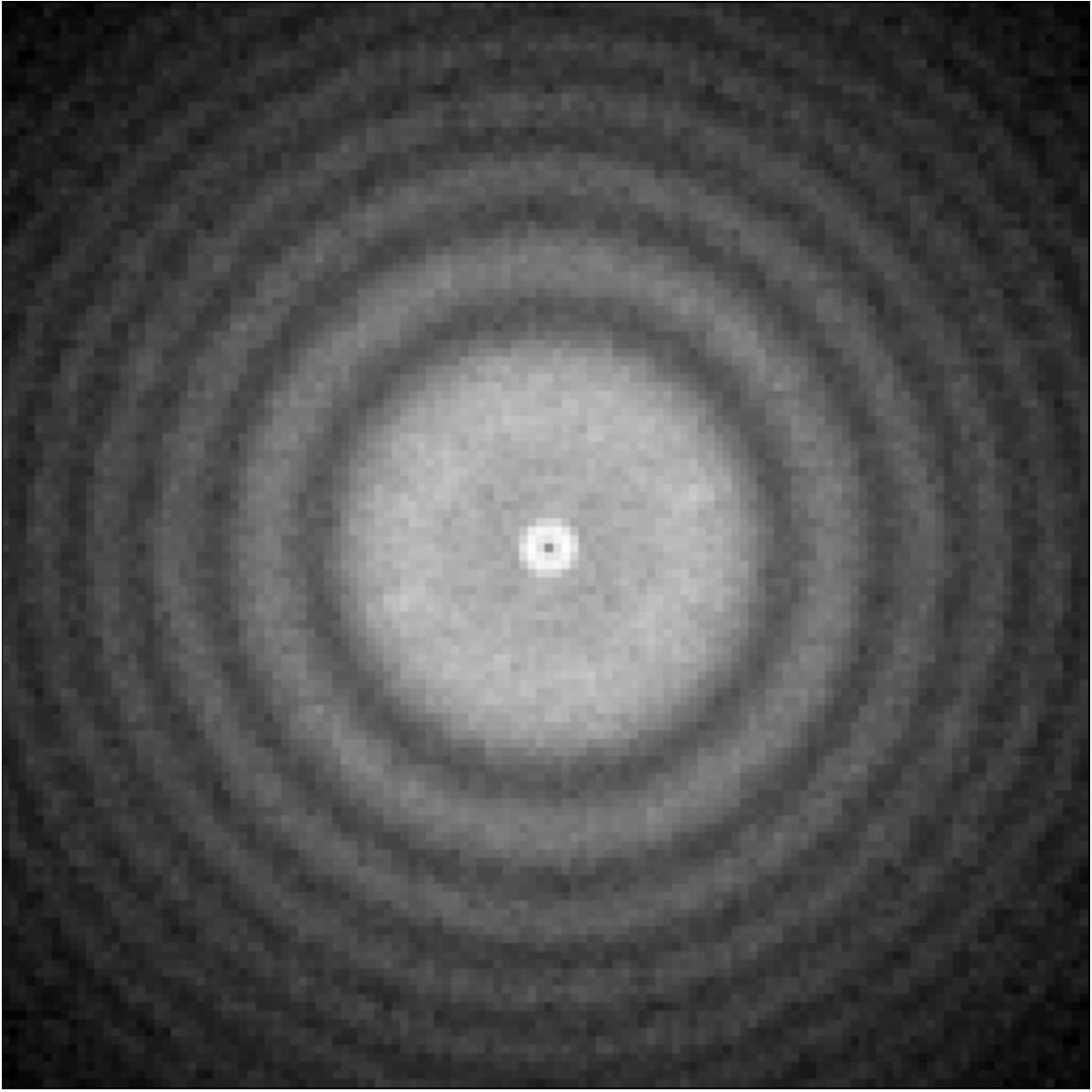}} &
{\includegraphics[width=0.22\linewidth]{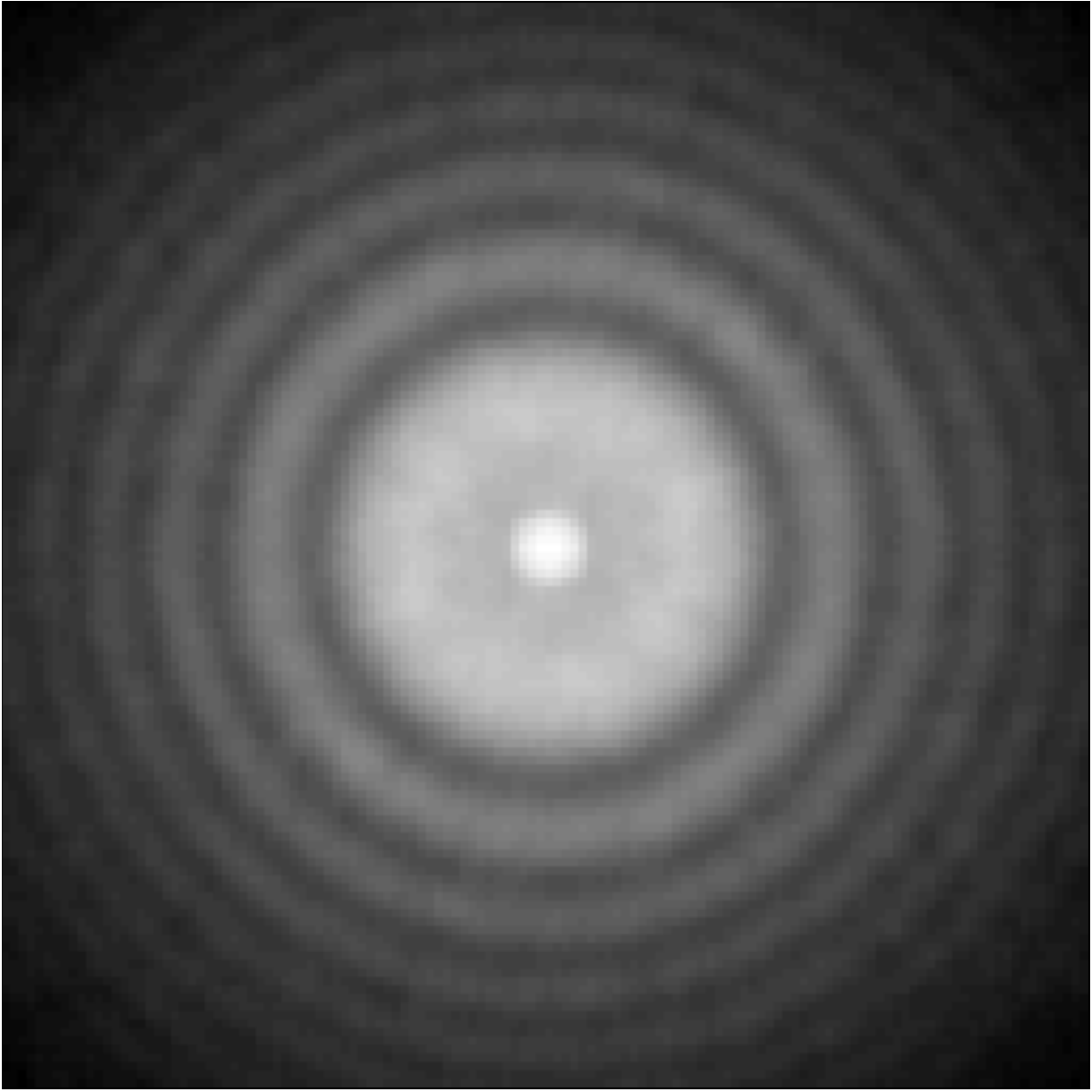}}\\
{\includegraphics[width=0.22\linewidth]{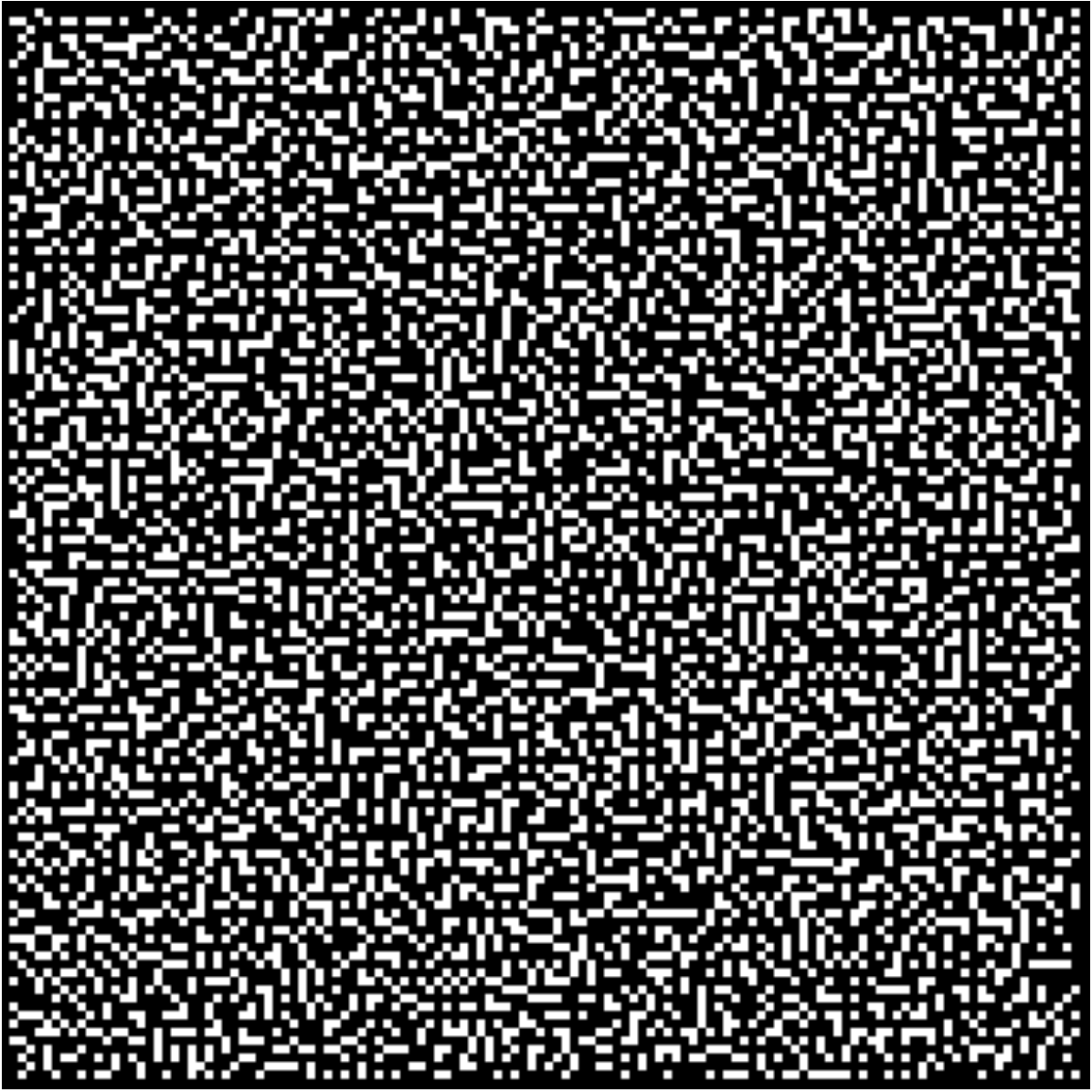}} &
{\includegraphics[width=0.22\linewidth]{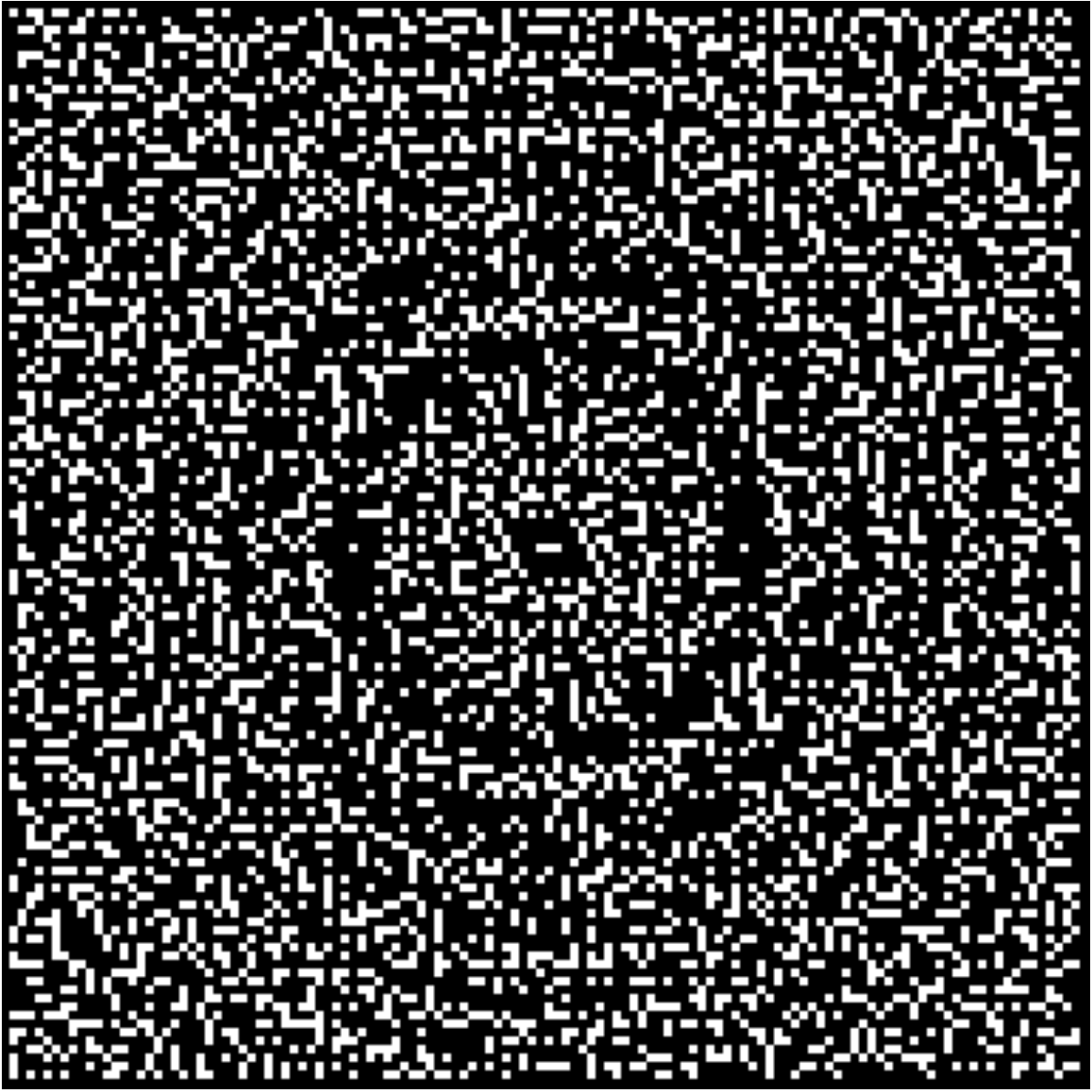}} &
{\includegraphics[width=0.22\linewidth]{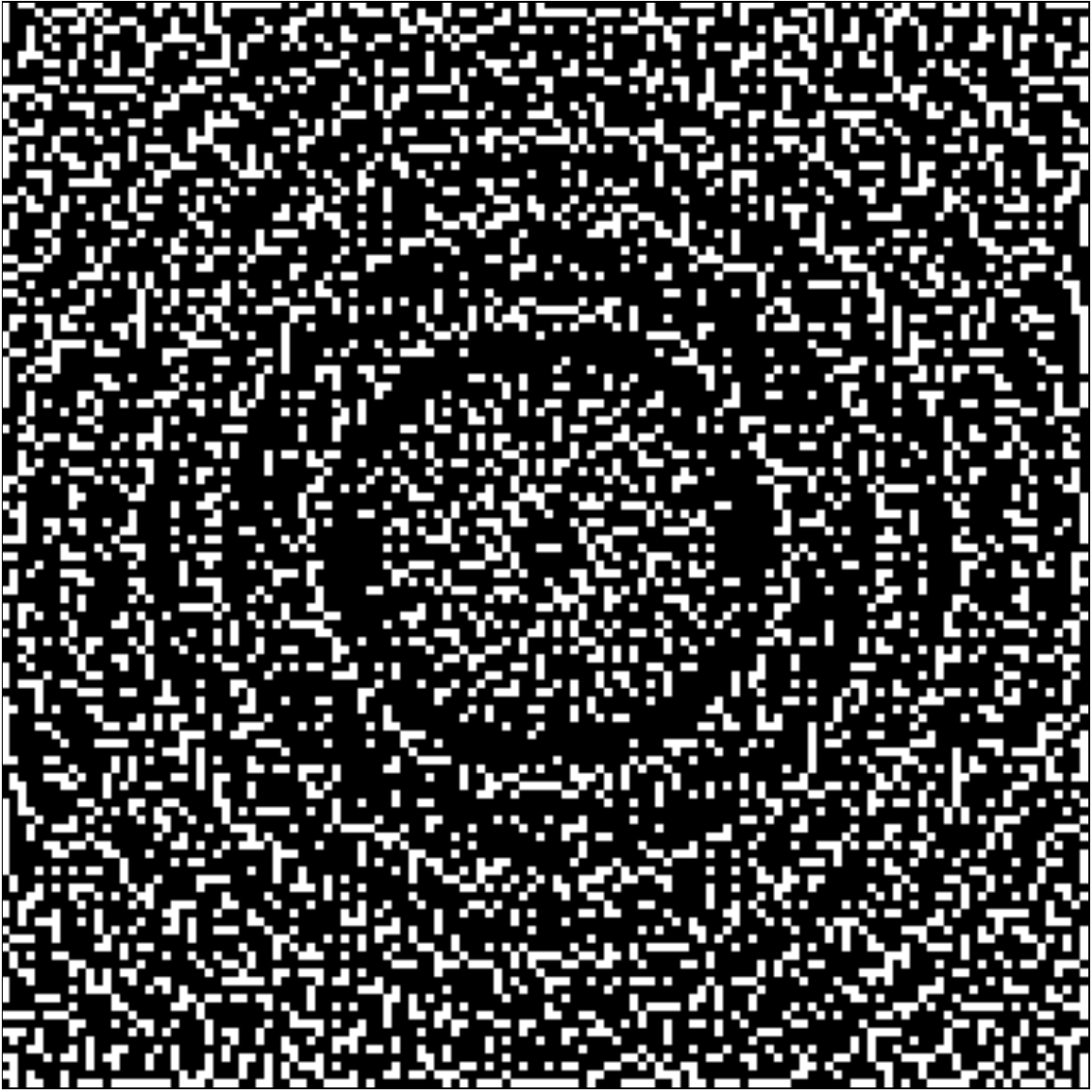}} &
{\includegraphics[width=0.22\linewidth]{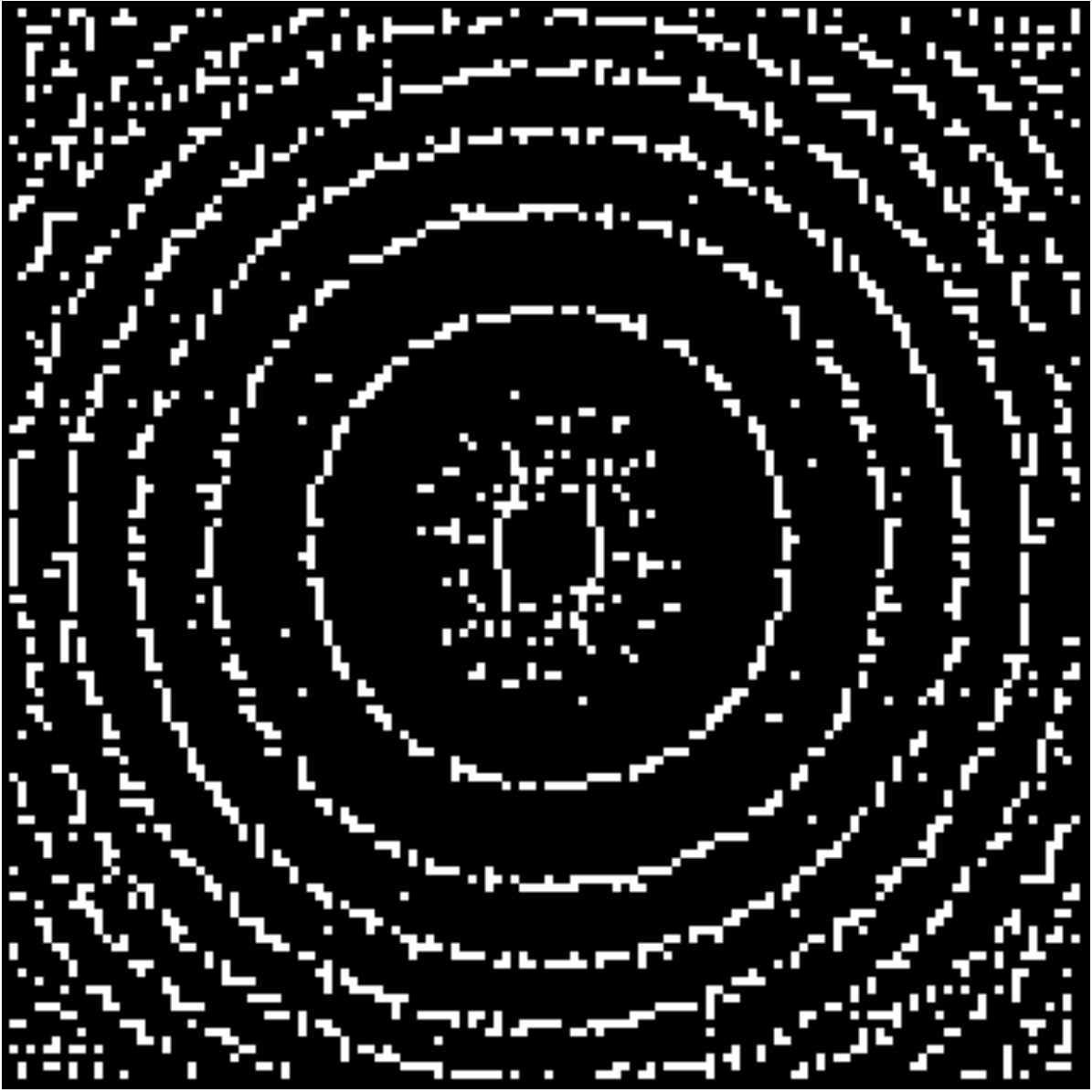}}
\end{tabular}
\caption{   
Power spectrum estimation and zero-crossings of the estimation for an 80S ribosome micrograph from the EMPIAR-10028 dataset \citep{wong2014empiar}. In the top row we present several power spectrum estimates (plotted on a logarithmic scale). The estimates are produced using the Periodogram estimator (leftmost column), Bartlett's method (center-left column), Welch's method (center-right column) and the Multitaper method with $L=9$ data tapers (rightmost column). Below each estimate we present its zero-crossings (which are determined as specified in Section~\ref{subsec:zero_cross}). 
Blocks of size $512 \times 512$ were used.
Once again, the zero-crossings of the CTF are most easily identifiable in the multitaper estimates.}
\label{fig:compare2}
\end{figure*}

 Figs. \ref{fig:compare}-\ref{fig:compare2}  present a comparison between $\hpsdper_{\y}$, $\hpsdbar_\y$, $\hpsdwel_\y$, and $\hpsdmt_\y$.
The CTF oscillations are best resolved by the multitaper estimator $\hpsdmt_\y$.

\subsection{Background subtraction}
\label{sec:bk}

In this section we present a method for removing the background spectrum and further reducing the variability of the estimator $\hpsdmt_\y$. 
We do this by first estimating the radial profile of the background spectrum and removing this estimation from $\hpsdmt_\y$.
Further, we show that the set of squared CTFs is contained in the span of a steerable basis and project our estimate onto that basis.

\subsubsection{Estimating the  background }
\label{subsec:bksub1}

We saw in~\eqref{equ:true_psd} that the micrograph power spectrum can be expressed as the sum of two spectra: the clean, filtered power spectrum $\vert \ctf(\vg) \vert^2\,\psd_\x(\vg)$ and the background  $\psd_\e$. 
The background-subtracted power spectrum is therefore
\begin{equation}
\psd_{\y} (\vg) - \psd_{\e} (\vg) = \vert \ctf (\vg) \vert^2 \, \psd_{\x} (\vg).
\end{equation}
An estimate of $\psd_\y - \psd_\e$ is used by many methods to estimate the CTF parameters $\phi$ \citep{rohou2015ctffind4, zhu19977gaussian}.
Their success therefore depends on accurate estimation of the background $\psd_\e$.

The background is influenced by many factors and accurately modeling these factors is an open challenge.
Many background estimation methods instead treat the background as a radially symmetric and slowly varying function %that is monotonically decreasing with increasing radial frequency
\citep{frank2996book}. 

The background estimation problem can be formulated as a curve fitting problem \citep{frank2996book}. We note that the background should coincide with $\hpsdmt_\y$ at the zero-crossings of $\ctf$. Furthermore, the background should be strictly smaller than $\hpsdmt_\y$ at spatial frequencies where $\ctf$ does not have a zero-crossing (since $\psd_\x$ is strictly  positive). We therefore estimate the background by minimizing the difference between $\hpsdmt_\y$  and $\psd_\e$.

While the radial profile of the background is monotonically decreasing in most settings,  this is not the case when a Gatan K2 direct detector is used in counting mode with a high dose rate. Rather, the background will be monotonically decreasing in the lower frequencies and monotonically increasing in higher frequencies \citep{li2013background}. Since a monotonically decreasing function, as well as a function that is at first monotonically decreasing and later monotonically increasing, must be convex, we model the background as the non-negative, convex function that is closest to, and no larger than, $\hpsdmt_\y$.

We propose estimating the background $\psd_\e$ through LP.
Specifically, we minimize the $\ell_1$ norm of the background-subtracted power spectrum estimate subject to several linear constraints.
The first constraint ensures that the background-subtracted power spectrum estimate is non-negative, while the remaining constraints ensure that $\hpsd_\e$ is a non-negative and convex. 

Since we assume the background is radially symmetric, we consider its radial profile. To this end, we calculate the radial average of $\hpsdmt_\y(\vg)$, which we denote, by a slight abuse of notation, $\hpsdmt_\y(r)$. The radial averaging is performed by projecting $\hpsdmt_\y(\vg)$ on the circularly symmetric (i.e., purely radial) elements of a steerable basis (see Section \ref{subset:steerable}). 

The resulting linear program, whose result we denote by $\hpsdblp$, is then
\begin{equation}
\begin{array}{ll}
\displaystyle \operatorname*{minimize}_{\hpsdb} &  \underset{r=0,\frac{1}{K},\dots,\frac{m}{K}}{\sum} {\hpsdmt_{\y} \left( r \right)} - \hpsdb\left( r \right)\\
 \text{subject to}  &
\hpsdb ( r ) \leq {\hpsdmt_{\y} \left( r \right)}, \quad r=0,\dots,\frac{m}{K}\\
 & \hpsdb \left( r + 1\right)  + \hpsdb \left( r -1\right)  \ge 2 \hpsdb \left( r \right), r=1,\dots,\frac{m}{K}\\
 &  \hpsdb \left( r \right) \ge 0,  \quad r=0,\dots,\frac{m}{K},
\end{array}
\end{equation}
where 
$\hpsdb = \begin{bmatrix} \hpsd(0), \ldots, \hpsd(m/K)) \end{bmatrix}^T$, and 
  $0<m/K \le 0.5$  is the  spatial frequency above which $\hpsdmt_{\y} \left( r \right)$ is typically dominated by noise.  As its default, ASPIRE-CTF sets $m/K=3/8$.
 
\begin{figure}[t]
\centering
\subfigure[] {\label{fig:back_est} \includegraphics[width=0.44\linewidth]{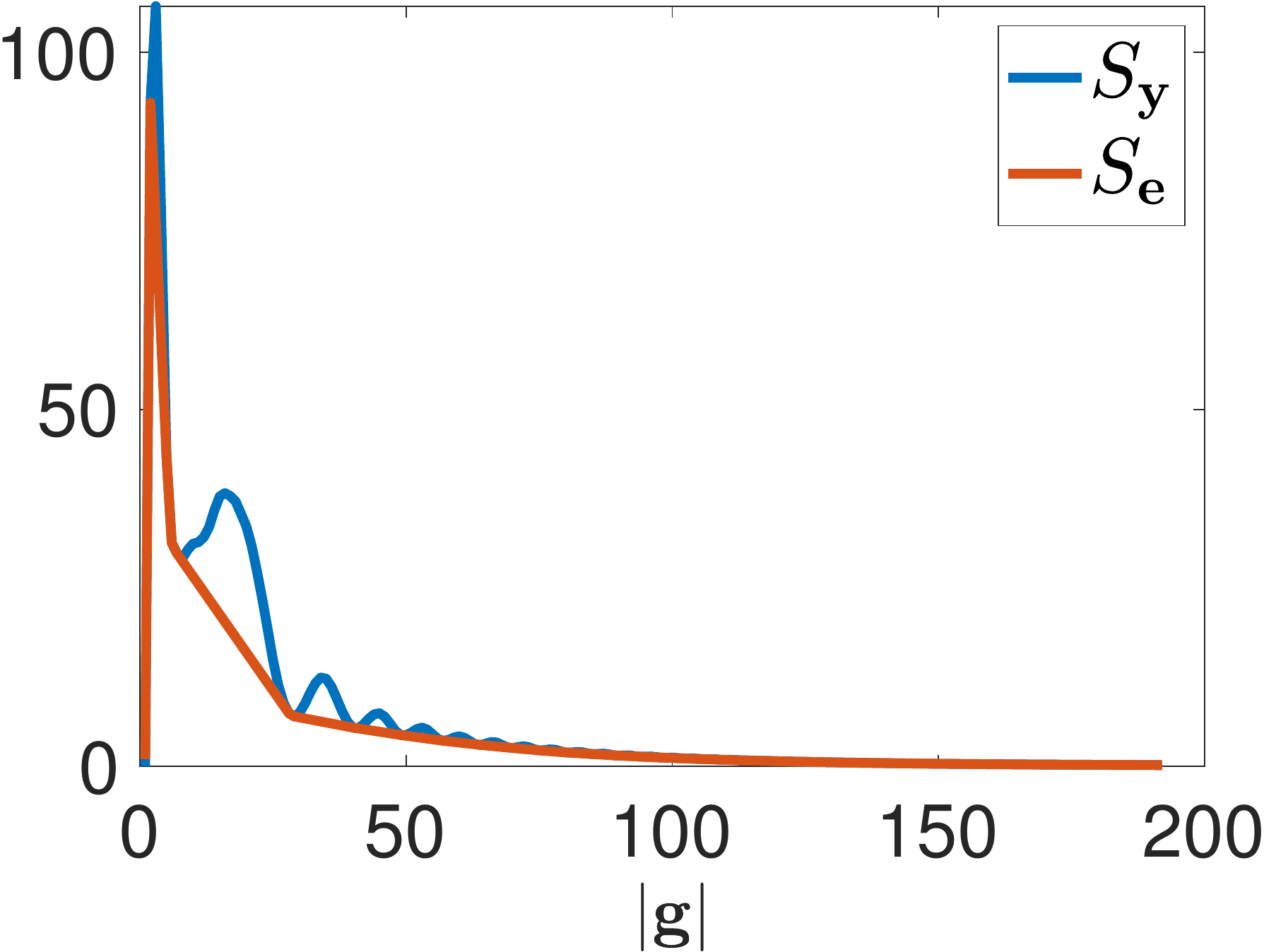}}
\hspace{0.3cm}
\subfigure[] {\label{fig:back_sub} \includegraphics[width=0.44\linewidth]{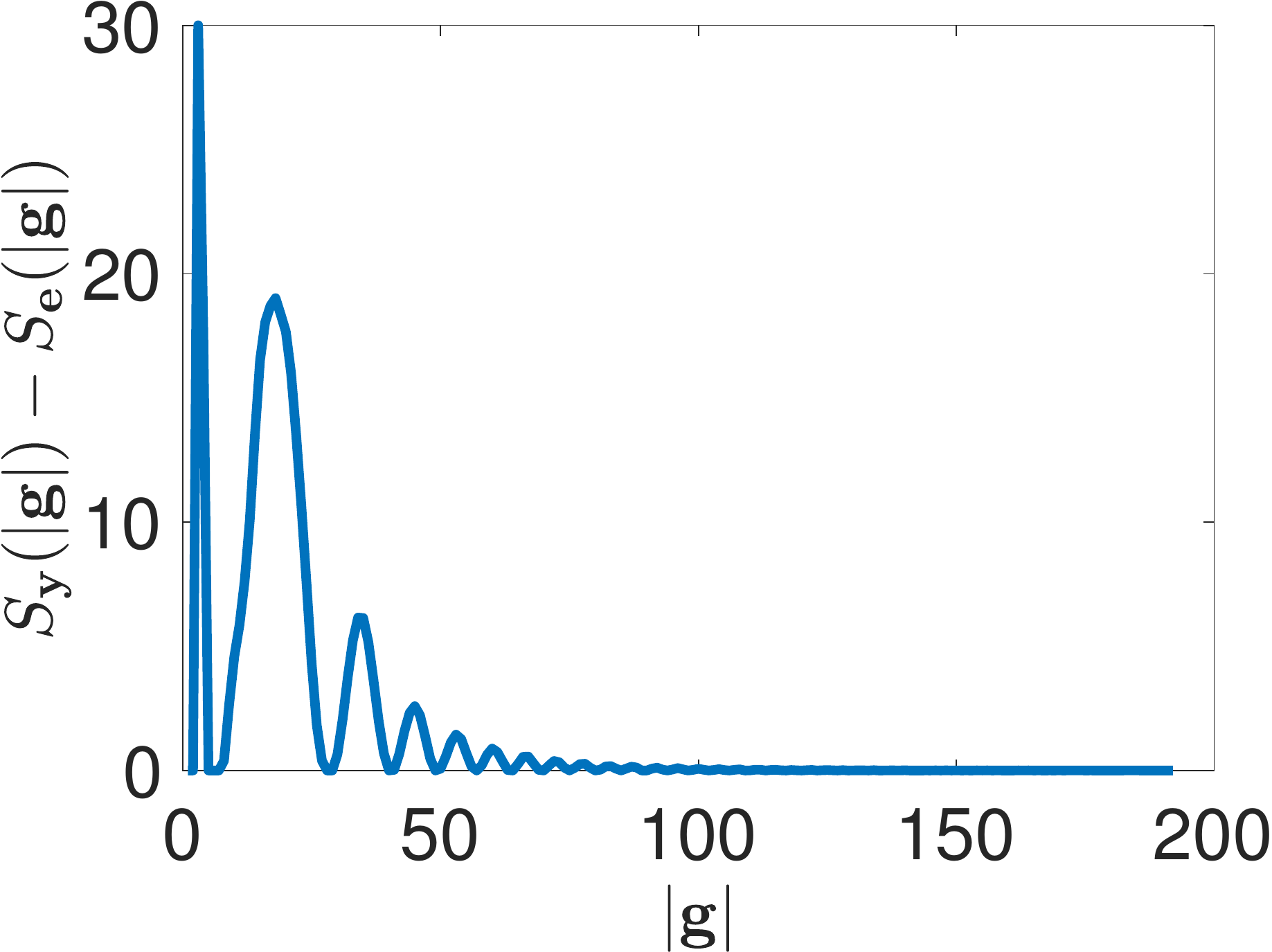}}
\caption{Background estimation for a $\beta$-galactosidase micrograph from the EMPIAR-10017 dataset \citep{scheres2015semi}.
On the left is the 1D radial profile of the multitaper power spectrum estimate $\hpsdmt_{\y}\left(r\right)$ and the estimated background $\hpsdblp \left(r \right)$.
On the right is the background-subtracted power spectrum.}
\label{fig:background}
\end{figure}

 We present the result of our linear program in Fig. \ref{fig:background}. 
Expanding the 1D background spectrum to a 2D function, we again abuse notation slightly and set $\hpsdblp (\vg) = \hpsdblp (r)$ for all $\vg \in [-1/2, 1/2]^2$.
We denote the background-subtracted power spectrum estimate $\hpsdmt_\y(\vg)-\hpsdblp(\vg)$ by $\hpsdlp_\z(\vg)$ where $\z = h_{\phi} * \x$. 
 
 A different LP that can be used for background estimation was suggested in \citep{huang2003env}. However, contrary to our non-parametric approach which assumes convexity alone,  \citep{huang2003env} suggests a LP based on parametric estimation.

\subsubsection{Expansion over a steerable basis}
\label{subset:steerable}

In this section, we show that any function of the form~\eqref{equ:ctf_sine}-\eqref{equ:astigmatism}  is contained in a low-dimensional subspace spanned by a set of steerable basis functions, such as a Fourier--Bessel basis \citep{zhao2013fourier,Zhao2016fb} or prolate spheroidal wave functions (PSWFs) \citep{landa2017prolates, landa2018prolates}.  
We will use this property to further reduce the variability of the power spectrum estimator by projecting the background-subtracted power spectrum estimate $\hpsdlp_\z(\vg)$ onto this subspace.

A steerable basis consists of functions $f_{k,q}(\vgmag) \, \euler^{\imunit k \vgang}$, where $k \in \Integer$ and $q = 0, \ldots, p_k-1$ for some $p_k \ge 0$.
The radial part $f_{k,q}(r)$ depends on the specific choice of basis (e.g., in a Fourier--Bessel basis, it is a scaled Bessel function of order $q$) and does not enter explicitly into our analysis.
We shall therefore leave it unspecified.
A given function in polar coordinates may be decomposed in the basis as
\begin{equation}
\label{equ:steerable}
x(\vgmag, \vgang) = \sum_{k=-\infty}^\infty \sum_{q=0}^{\infty} a_{k,q} \, f_{k,q}(\vgmag) \, \euler^{\imunit k \vgang},
\end{equation}
where $a_{k,q} \in \Complex$ is the coefficient corresponding to angular frequency $k$ and radial frequency $q$.

To determine the steerable basis expansion of the CTF~\eqref{equ:ctf_sine}, we consider its Taylor expansion around $\dfone - \dftwo = 0$,
\begin{multline}
\ctf ( \vg ) = \sum_{\substack{n = 0 \\ n~\text{even}}}^{P} \frac{\left( -1 \right)^{\frac{n}{2}+1}}{n!} \sin ( \phase^0 ( \vgmag )) C_{n,\phi}(\vg) \\+  \sum_{\substack{n=1 \\ n~\text{odd}}}^{P} \frac{\left( -1 \right)^{\frac{n+1}{2}}}{n!} \cos ( \phase^0 ( \vgmag )) C_{n,\phi}(\vg) + R_P(\vg),
\end{multline}
where $R_P(\vg)$ is the remainder term,
\begin{equation*}
C_{n,\phi}(\vg) = \left( \frac{1}{2}\pi \lambda \left( \dfone - \dftwo \right) \cos \left( 2 ( \vgang - \dfang) \right) \frac{ \vgmag ^2}{p^2} \right)^n,
\end{equation*}
and
\begin{equation*}
\phase^0 ( \vgmag ) = \frac{1}{2} \pi \lambda  \vgmag^2 (\dfone+\dftwo) - \frac{1}{2} \pi \lambda^3  \vgmag^4 C_s + w
\end{equation*}
is the non-astigmatic phase function.

The remainder term is bounded by a function of
\begin{equation*}
\left(\frac{\dfone - \dftwo}{\dfone+\dftwo}\right)^{(P+1)}
\end{equation*}
and is therefore small when astigmatism is small, which is the case for experimental cryo-EM data. 
We therefore conclude that
\begin{equation}
\label{equ:ctf_approx}
\ctf(\vg) \approx -\sin(\phase^0 (\vgmag)) - \cos(\phase^0(\vgmag)) C_{1,\phi}(\vg)
\end{equation}
is a good approximation of the CTF.

Since $\cos(\alpha) = \frac{1}{2}(\euler^{-\imunit \alpha} + \euler^{-\imunit \alpha})$, we rewrite \eqref{equ:ctf_approx} as
\begin{multline}
\label{equ:taylor_ctf}
\ctf(\vg) \approx -\sin(\phase^0(\vgmag)) \\
-\frac{1}{4p^2} \cos(\phase^0(\vgmag)) \pi \lambda (\dfone-\dftwo) \euler^{-\imunit 2\dfang} r^2 \, \euler^{\imunit 2\vgang} \\
-\frac{1}{4p^2} \cos(\phase^0(\vgmag)) \pi \lambda (\dfone-\dftwo) \euler^{\imunit 2\dfang} r^2 \, \euler^{-\imunit 2\vgang}.
\end{multline}
Comparing \eqref{equ:taylor_ctf} and \eqref{equ:steerable}, we see that only terms corresponding to $k = -2$, $0$, and $2$ are present. 

Concretely, we compute the coefficients $a_{k,q}$ of the expansion of $\sqrt{\hpsdlp_\z}$ over the steerable basis functions with radial frequencies to $k = 0, \pm 2$. The coefficients are computed through an inner product on a $K \times K$ grid:
\begin{equation}
a_{k,q} = \frac{1}{K^2} \sum_{\vg \in M_K} \sqrt{\hpsdlp_\z(\vg)} f_{k, q}(\vgmag) \euler^{\imunit k \vgang},
\end{equation}
where $r$ and $\vgang$ are the polar coordinates of $\vg$.
Evaluating \eqref{equ:steerable} for these $a_{k,q}$ and squaring the result then gives a new power spectrum estimate, which we denote as $\hpsd_\z$.

\begin{figure}[t]
\centering
\subfigure[]{\label{subfig:psd}\includegraphics[width=0.3\linewidth]{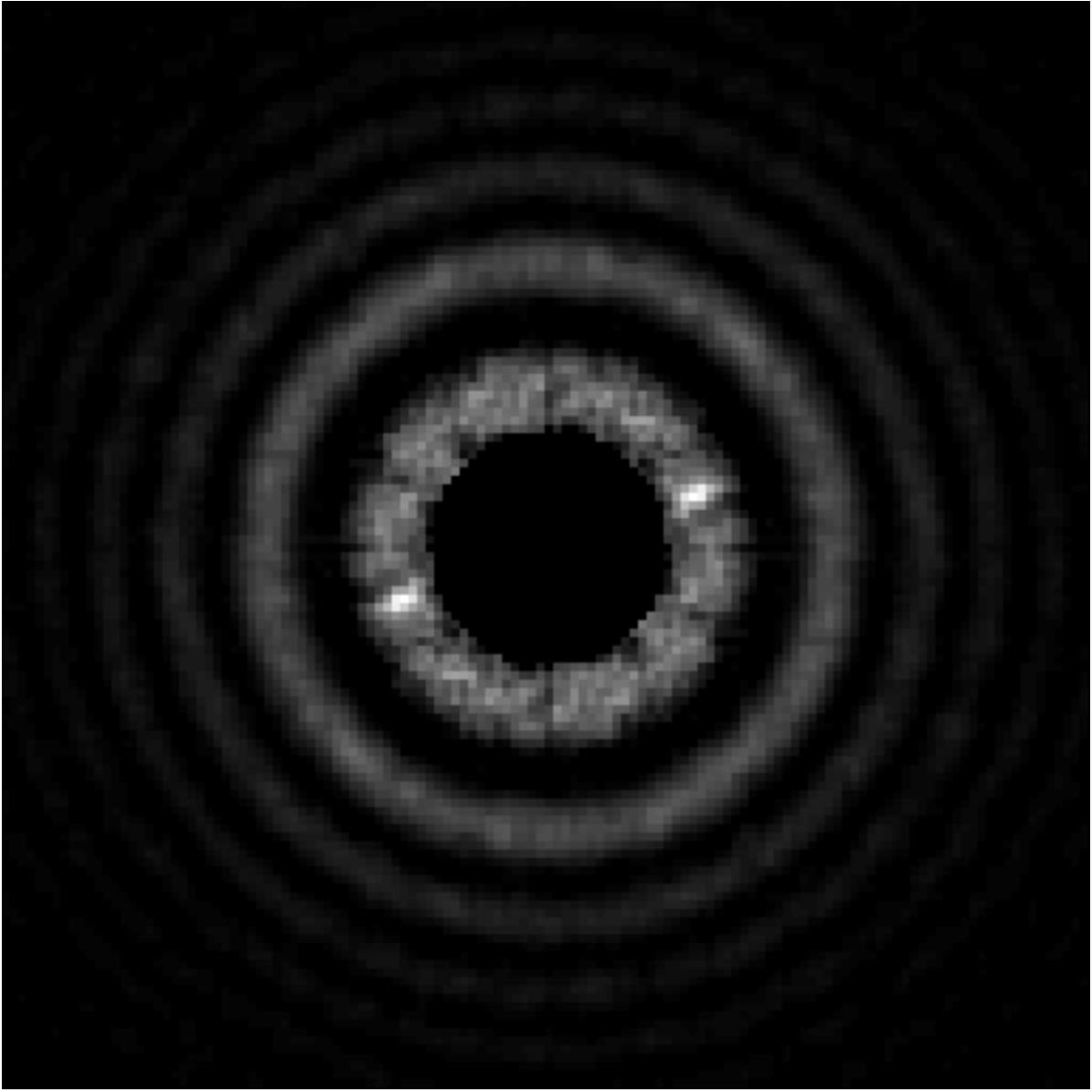}}
\hspace{0.1cm}
\subfigure[]{\includegraphics[width=0.3\linewidth]{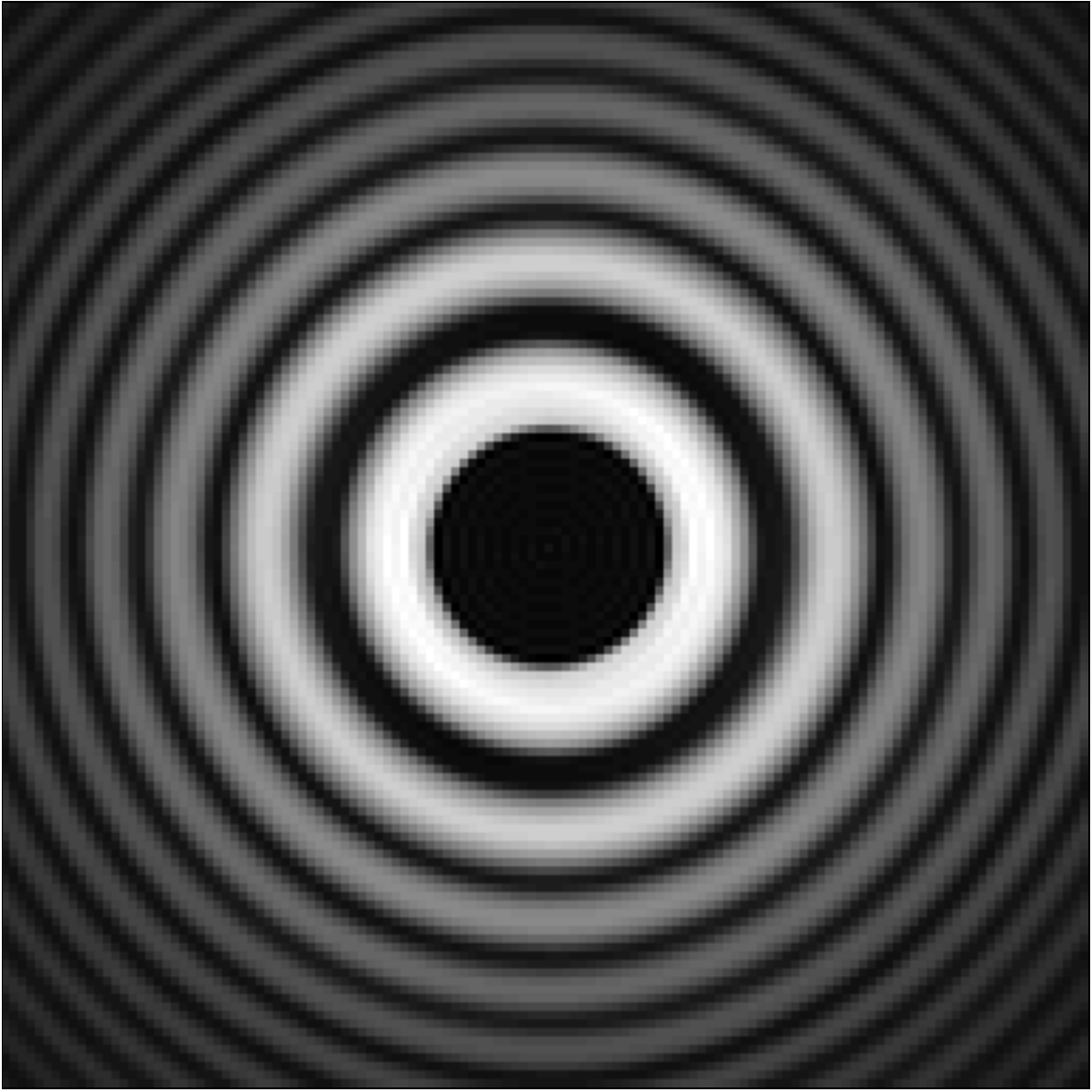}}
\hspace{0.1cm}
\subfigure[]{\includegraphics[width=0.3\linewidth]{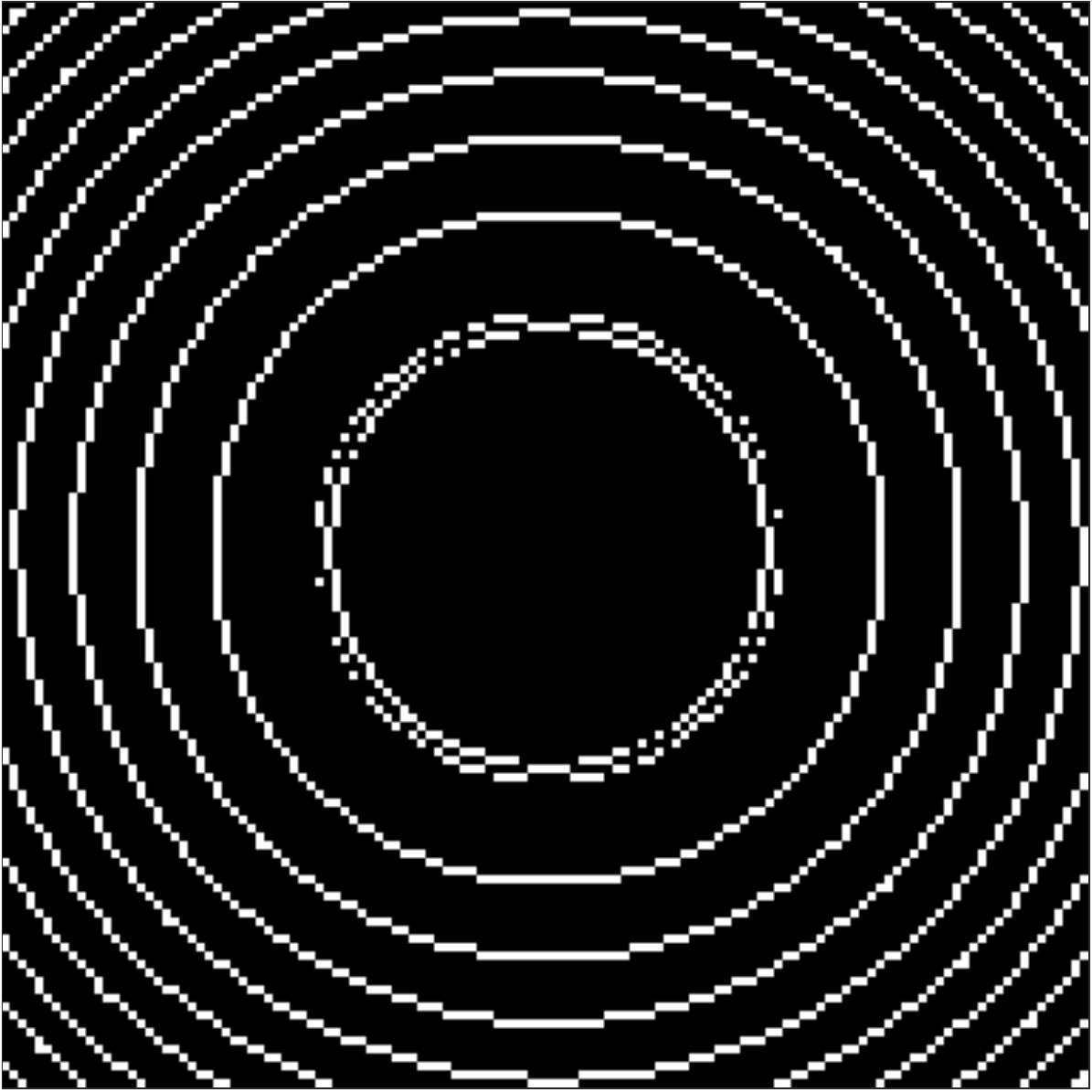}}
\caption{Power spectral density estimates. (a) Background-subtracted estimate $\hpsdlp_\z$. (b) Projection onto steerable basis functions for $k = 0, \pm 2$. (c) Zero-crossings of panel b, determined as specified in Section~\ref{subsec:zero_cross}.}
\label{fig:PSD_complete}
\end{figure}

In Fig. \ref{fig:PSD_complete}, we present the results of our power spectrum estimation method on a micrograph from the EMPIAR-10028 dataset \citet{wong2014empiar}.
This includes multitaper estimate $\hpsdmt_\y$ as well as the background-subtracted estimate $\hpsdlp_\z$ and the projection onto the steerable basis with $k = 0, \pm 2$. 
The result is smooth enough that many of the zero-crossings of the power spectrum can be easily resolved.

\subsection{CTF Parameter Estimation}

In the previous sections we have introduced our method for estimating the background-subtracted power spectrum.
In this section we will discuss two methods that use this estimate to recover the defocus and astigmatism of the micrograph.

\subsubsection{CTF estimation through correlation}
\label{subsec:cross_corr}

The power spectrum $\psd_\x$ is a slowly-varying function of radial frequency. It follows that the oscillations in $|\ctf(\vg)|^2 \, \psd_\x(\vg)$ are due to those of $|\ctf (\vg)|^2$. 
As a consequence, the square root of the background-subtracted power spectrum $|\ctf(\vg)| \, \psd_\x^{1/2}(\vg)$ is proportional to the absolute value of the CTF (we note that using the square root instead of the actual power spectrum estimate reduces the influence of large values). It follows that the correlation test is a useful tool in estimating CTF parameters. Indeed, 
many CTF estimation methods therefore estimate the defocus parameters $\phi$ through correlation with a simulated CTF magnitude \citep{rohou2015ctffind4,mindell2003ctffind3,kai2016gctf}.

The Pearson correlation of $|\ctf(\vg)|$ and $\hpsd_\z^{1/2}(\vg)$ is
\begin{equation}
\label{equ:corr_def}
P_{cc}(\phi) = \frac{\sum_{\vg \in R} |\ctf(\vg)| \, \hpsd_\z^{1/2}(\vg)}{\left( \sum_{\vg \in R} |\ctf(\vg)|^2 \, \sum_{\vg \in R} \hpsd_\z(\vg) \right)^{1/2} },
\end{equation}
where $R$ is the set of frequencies over which correlation is computed, and will be defined below. 
To optimize $P_{cc}(\phi)$, we first need an initial guess for the parameters $\phi$.
For this, we follow \citet{kai2016gctf} and first consider non-astigmatic CTFs where $\dfone = \dftwo$, which renders the value of $\dfang$ irrelevant.
We thus calculate $P_{cc}(\phi)$ for $\phi = (\df, \df, 0)$ with $\df$ on a 1D grid from $\df_{\mathrm{min}}$ to $\df_{\mathrm{max}}$ with a step of $\df_{\mathrm{step}}$.
Since $\ctf$ is considered (at this stage) to be radially symmetric, we define the set of frequencies over which correlation is computed as 
\begin{equation}
R = \left\{ m_1, m_1+ \frac{2}{N}, \ldots, \frac{3}{8} \right\} \times \{0\},
\end{equation} 
where $m_1$ is the first maximum of the radial profile of $\hpsdlp_\z$. That is, we only consider frequencies $\vg$ along a 1D radial profile and, furthermore, ignore the very low and very high frequencies (since the very low frequencies may dominate the cross-correlation result and the very high frequencies are strongly effected by the envelope function). The $\df$ which maximizes $P_{cc}(\phi)$ on this grid is denoted $\df_\star$.

To estimate the astigmatism of the CTF, we compute the principal directions of the second-order moments of $\hpsd_\z^{1/2}$.
Specifically, we form the $2 \times 2$ matrix $\M$ given by
\begin{multline}
\begin{aligned}
& M_{1,1} = {\textstyle \sum_{\vg \in M_K}} \, g_1^2  \, \hpsd_\z^{1/2}(\vg),\\
& M_{1,2} = M_{2,1} = {\textstyle \sum_{\vg \in M_K}} \, g_{1} g_{2} \, \hpsd_\z^{1/2}(\vg),\\
& M_{2,2} = {\textstyle \sum_{\vg \in M_K}} g_2^2 \, \hpsd_\z^{1/2}(\vg).
\end{aligned}
\end{multline}
The eigenvalues $\mu_1$ and $\mu_2$ of $\M$ estimate the size of the major and minor axes in $\hpsd_\z^{1/2}$.
We therefore expect the ratio $\mu_1/\mu_2$ to approximate $\dfone/\dftwo$.
Combining this with the estimated mean defocus $\df_\star$, we get
\begin{align}
\frac{1}{2}(\dfone + \dftwo) &= \df_\star \\
\frac{\dfone}{\dftwo} &= \frac{\mu_1}{\mu_2},
\end{align}
which has the solution
\begin{align}
\df_{1,\star} &= \frac{2\mu_1}{\mu_1+\mu_2} \df_\star, \quad
\df_{2,\star} &= \frac{2\mu_2}{\mu_1+\mu_2} \df_\star.
\end{align}

In order to improve our estimation of the defocus parameters, we run gradient descent on  $P_{cc}(\phi)$. As we no longer approximate the image as non-astigmatic, we define the set of frequencies over which correlation is computed as $R = M_K$.  We now initialize our gradient descent at $\phi = (\df_{1,\star}, \df_{2,\star}, \dfang)$, where $\dfang$ is set as detailed in \citep{kai2016gctf} to an arbitrarily selected $0\le a < \pi/6$ (e.g. $a=\pi/12$) and $(a+\pi/6)$, $(a-\pi/6)$, $(a+\pi/3)$, $(a-\pi/3)$ or $(a-\pi/2)$.
One run of gradient descent is performed for each value of $\dfang$ and the result with the highest value of $P_{cc}(\phi)$ is kept.
The resulting $\phi$ is our defocus estimate for the micrograph.

We note that, as is done in \citep{rohou2015ctffind4,mindell2003ctffind3,kai2016gctf}, we discard information in the lower and higher frequencies of $ \hpsd_\z$. These frequencies can be determined by the user. As default values we use those suggested in \citep{rohou2015ctffind4}.

\subsubsection{CTF estimation through zero-crossings}
\label{subsec:zero_cross}

We have seen in the previous sections that the true background-subtracted power spectrum is $\vert \ctf (\vg) \vert^2 \psd_\x (\vg)$.
Under the assumption that $\psd_\x (\vg)$ is slowly-varying, it follows that at any frequency where the background-subtracted power spectrum reaches a minimal value of zero, the CTF must reach a zero-crossing. 

Furthermore, even if the aforementioned assumption did not hold true, we could still infer the frequencies where the CTF reaches a zero-crossing.
This is due to the fact that the zero-crossings of the CTF are known to create concentric, nearly elliptical rings, centered around the origin (see Section~\ref{sec:formation}).
Therefore, this can be used a cue to differentiate between any minima of $\vert \ctf (\vg) \vert^2 \, \psd_\x (\vg)$ that stem from the zero-crossings of $\ctf$ and any minima that stem from $\psd_\x$.

As we show in Fig.~\ref{fig:PSD_complete}(c), our estimation of the background-subtracted power spectrum enables easy detection of several elliptical rings where $\hpsd_\z$ reaches its minima. To do this, we define any pixel with a value smaller than that of at least six out of its eight neighbors as a zero-crossing.
Once the minima of $\hpsd_\z$ are found, we discard any frequency that is not on a closed ring. Furthermore, we verify that the spatial frequencies of pixels residing on closed rings representing the minima of $\hpsd_\z$ form ellipses approximately centered at the origin.
We are then left with frequencies of several zero-crossings of the CTF.

Since we have $\ctf(\vg) = - \sin(\phase(\vg))$, we reach a zero-crossing of the CTF when $\phase(\vg)$ is an integer multiple of $\pi$.
Formally, the set of spatial frequencies on the $\ell$th ring of zero-crossings, denoted by  $G_\ell$, satisfies
\begin{equation}
\phase(\vg_\ell) = \pi \ell.
\label{equ:zero_constraint}
\end{equation}
Empirically, we are typically able to identify at least three rings, that is, three different values of $\ell$.

Combining \eqref{equ:zero_constraint} for all $\vg$ in $\hat{G}_\ell$ and combining these for different values of $\ell$, we obtain an overdetermined system of equations.
Solving it yields an estimate for the defocus vector $\phi$.
To solve the system, we use the \textsf{trust-region-dogleg} method implemented in MATLAB (a variant of \citep{powel1970fsolve}).

 We note that the estimated positions of CTF zeroes are extremely sensitive to the method of background subtraction. This method should therefore be used to obtain an initial estimate of the defocus. Refinement of the solution can be done as in Section~\ref{subsec:cross_corr} using gradient-based optimization.

We note that typically both methods suggested in this section achieve similar results. However, while the zero-crossings-based method has lower computational complexity, the correlation-based method is more robust to noise. Therefore, for micrographs with very low SNR we recommend using the correlation-based method, while for cleaner micrographs we suggest using the zero-crossings method.

\section{Results}
\label{sec:experiment}

We present experimental results for the ASPIRE-CTF framework presented in this paper. We apply our framework to
datasets that are publicly available from the EMPIAR database~\citep{iudin2016empiar} or the CTF challenge \citep{marabini2015challenge}. Unless otherwise stated, in the experiments below we use $L=4$ tapers and project the power spectrum onto the steerable PSWF basis.

\subsection{Estimating CTF from movie frames}

The CTF may be estimated either from motion-corrected micrographs,  from several frames averaged in real space  or, alternatively, directly from the frames. This is done by estimating $\hpsdmt$ individually from each frame and averaging the estimates (see Fig.~\ref{fig:pipeline}). A benefit of estimating the CTF directly from the frames is that this practice enables us to correct for motion and estimate the CTF concurrently, thus speeding up the pipeline. Furthermore, any errors added by motion-correction will have no effect on the CTF estimation~\citep{bartesaghi2014frames}.

In this section, we compare the CTF estimates produced from motion-corrected micrographs to the estimation produced directly from the frames. We do this over several publicly available datasets, namely, EMPIAR-10002~\citep{bai2013empiar}, EMPIAR-10028~\citep{wong2014empiar}, EMPIAR-10242~\citep{zhang2019empiar}, and EMPIAR-10249~\citep{herzik2019dataset}. A summary of these datasets appears in Table~\ref{table:movies}.  

\begin{figure*}[t]
\begin{center}
\subfigure[]{\label{subfig:psd}\includegraphics[height=0.22\linewidth]{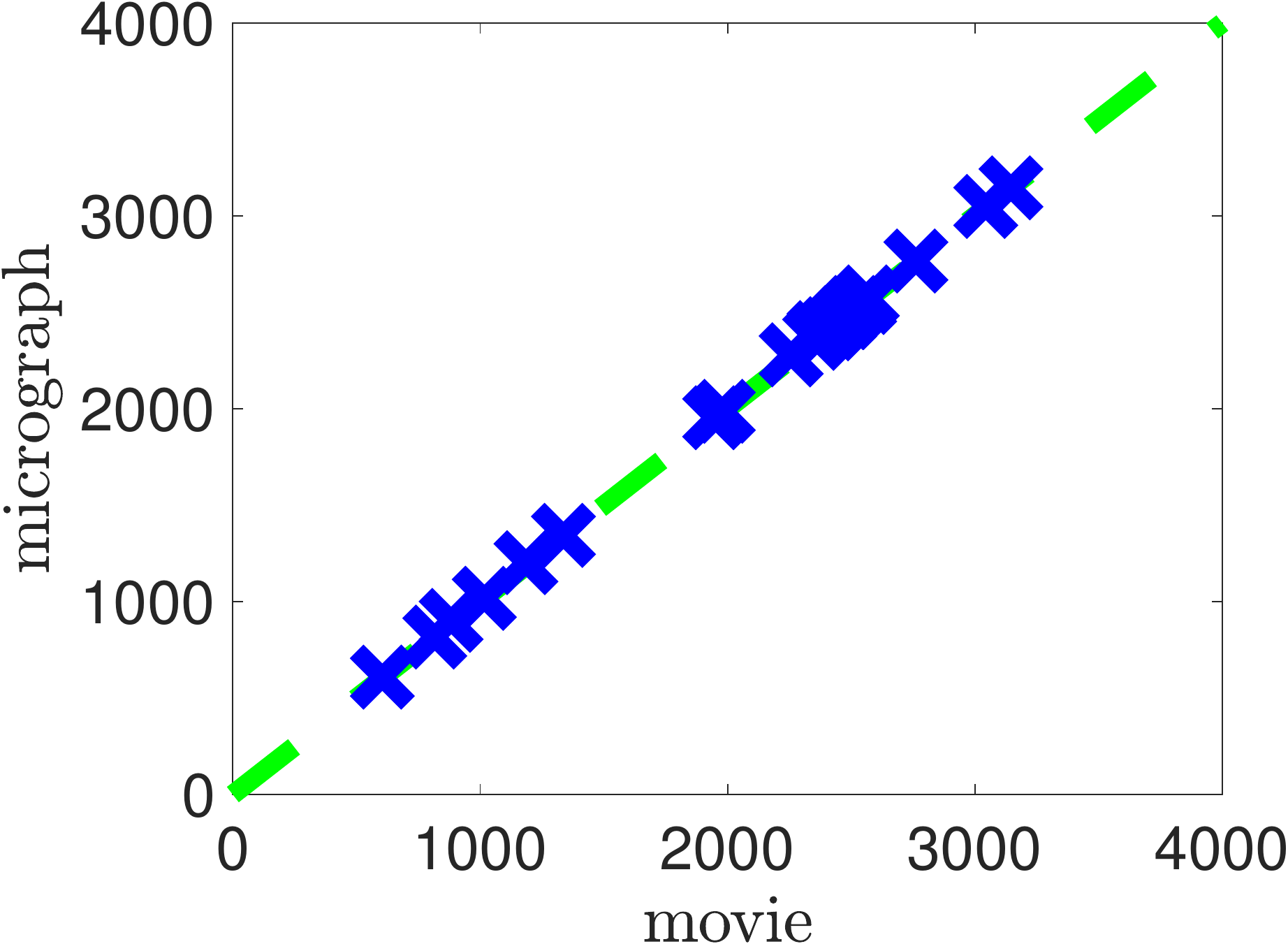}}
\subfigure[]{\label{subfig:psd}\includegraphics[height=0.22\linewidth]{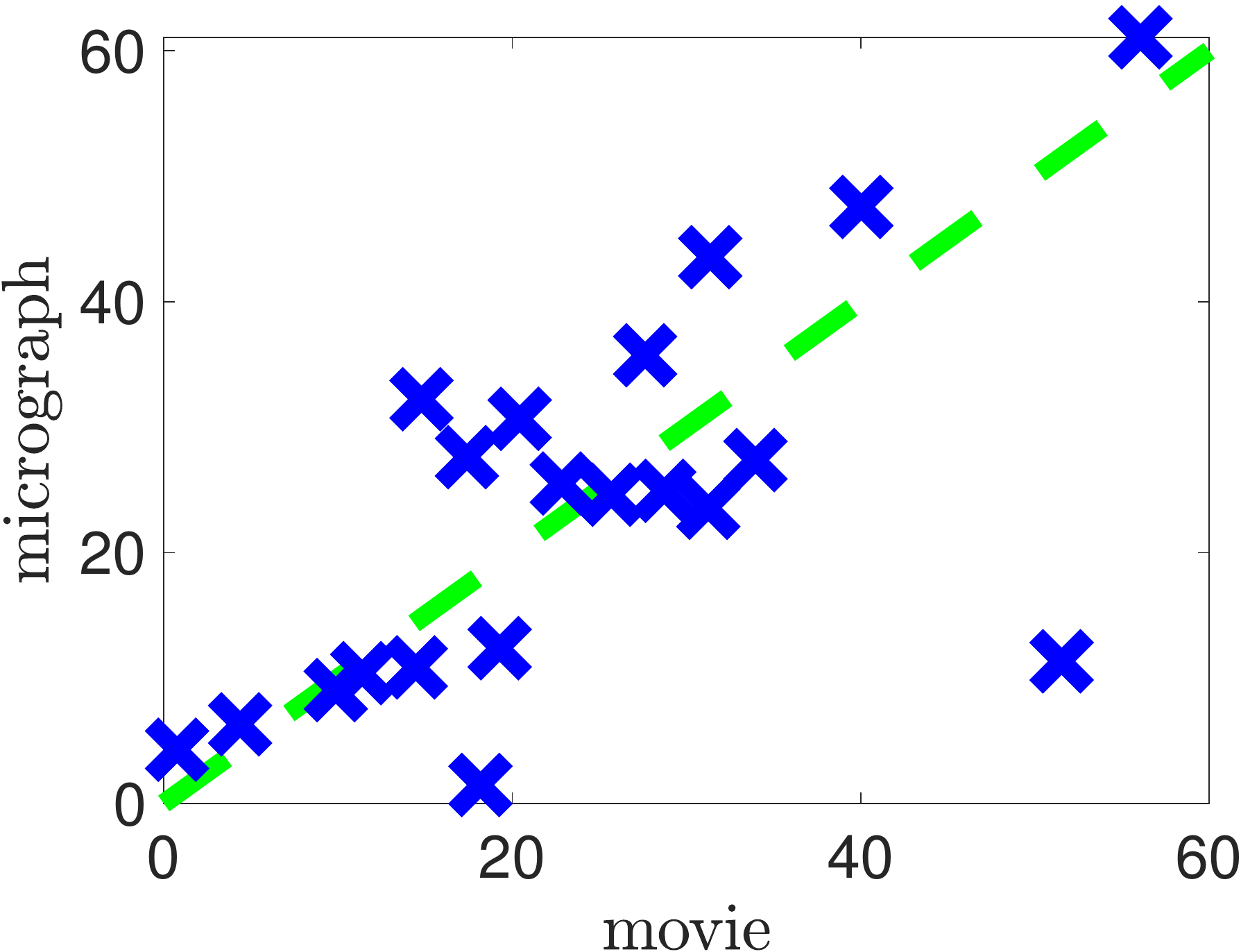}}
\subfigure[]{\label{subfig:psd}\includegraphics[height=0.22\linewidth]{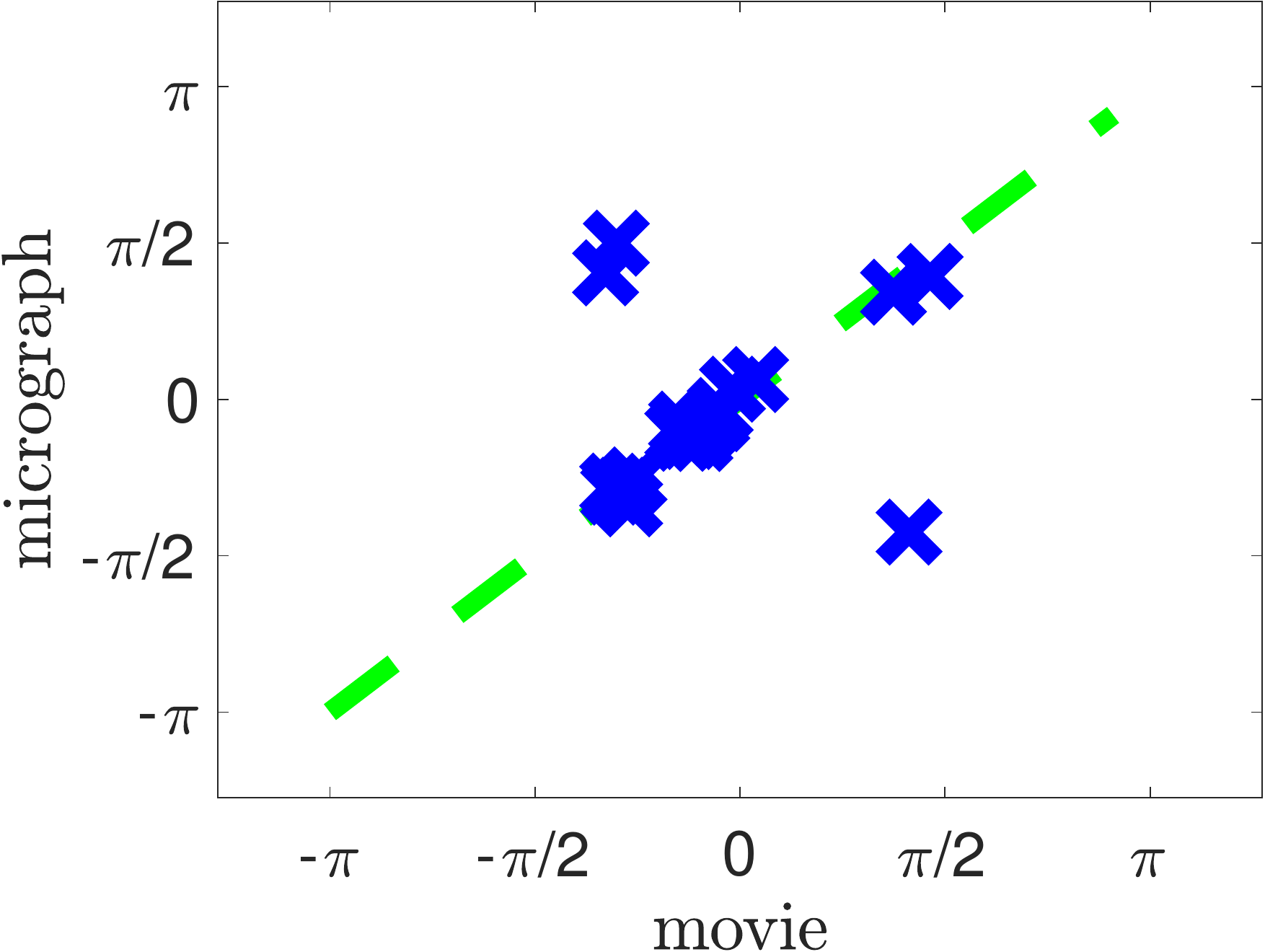}}
\caption{Defocus estimation on sample micrographs of the EMPIAR-10028, EMPIAR-10002, EMPIAR-10042 and EMPIAR-10049 datasets. We compare the estimation of each parameter when using the average power spectrum of the frames to the estimation when using the power spectrum of the motion-corrected micrograph. (a) Average defocus (in nm). (b) Astigmatism (in nm). (c) Angle $\alpha_f$ (in radians).}
\label{fig:movies}
\end{center}
\end{figure*}

While the EMPIAR-10028 and EMPIAR-10242 datasets contain both movies and motion-corrected micrographs, EMPIAR-10002 and EMPIAR-10249 contain movies alone. We therefore use MotionCor2 \citep{Zheng2017motion} to produce the motion-corrected micrograph for these two datasets. We present in Fig.~\ref{fig:movies}  a comparison of the astigmatism ($\dfone-\dftwo$), average defocus ($\dfone/2+\dftwo/2$), and astigmatism angle ($\alpha_f$) as estimated from a motion-corrected micrograph with an estimate produced from the raw movie frames. We note that, as expected, the parameters estimated from each of these methods are nearly identical.

\begin{table*}
\begin{center}
\begin{tabular}{ | c | c | c | c | c | c | c | }
  \hline			
 Dataset & Molecule & Pixel size (\AA) & Spherical & Voltage (kV) & Microscope & Detector \\
              &           &      & aberration &           &  & \\
 \hline
EMPIAR-10002 & 80S ribosome & 1.77 & 2.0 & 300& Polara & Falcon  \rom{1}\\
EMPIAR-10028 & 80S ribosome & 1.34 & 2.0 & 300 & Polara & Falcon  \rom{2}\\
EMPIAR-10242 & 2N3R tau filaments & 1.04 & 2.7 & 300 & Titan Krios & Gatan K2 Summit\\
EMPIAR-10249 & HLA dehydrogenase & 0.56 & 2.7 & 200 & Talos Arctica & Gatan K2 Summit\\
\hline
\end{tabular}
\end{center}
\caption{Description of the EMPIAR-10002, EMPIAR-10028, EMPIAR-10242 and EMPIAR-10249 datasets.}
\label{table:movies}
\end{table*}

\subsection{CTF Challenge}
\label{subsec:exp1}
The CTF challenge \citep{marabini2015challenge} consists of nearly $200$ micrographs of GroEL, 60S ribosome, apoferritin and TMV virus.
These micrographs are taken from eight experimental datasets and one synthetic dataset, each referred to by a  number ranging from $001$ to $009$. In the following, we restrict our attention to the experimental datasets, that is, sets $001$ through $008$.

The advantage of the CTF challenge is that each dataset is acquired using a different combination of microscope and camera, allowing for a qualitative comparison of CTF estimation methods for a variety of experimental setups.   
Notably, datasets $003$ and $004$ use a Gatan K2 direct detector in counting mode with a high electron dose, causing $\psd_\e$ to increase at high frequencies \citep{li2013background}.
Additionally, dataset $008$ has an especially low signal-to-noise ratio (SNR), rendering CTF estimation difficult. A summary of these datasets is presented in \citep{marabini2015challenge}.

The estimate of each micrograph's power spectrum is computed as detailed in Section~\ref{subsec:multi}. Specifically, we divide the micrograph into half-overlapping blocks of size $K \times K$, where $K=512$, and use $L=4$ tapers in the estimation. We then estimate the background spectrum as detailed in Section~\ref{subsec:bksub1} and expand the background-subtracted power spectrum over the PSWF basis in order to reduce variability in the power spectrum estimate (Section~\ref{subset:steerable}). We use the correlation-based method (Section~\ref{subsec:cross_corr}) to estimate defocus parameters, and denote the resulting vector of defocus parameters as $\phi_{a}^{(512)}$. 

In some cases, a power spectrum of size $512 \times 512$ may not capture the oscillations of the power spectrum with sufficient accuracy \citep{rohou2015ctffind4}. We therefore compute a second estimate of the power spectrum using half-overlapping blocks of size $1024 \times 1024$. As this reduces the number of blocks, the variance of the estimator will grow. We therefore use $L=16$ data tapers in this case. We use this estimate of the power spectrum to estimate a vector of  defocus parameters which we denote by $\phi_{a}^{(1024)}$. 

We compare our results to the estimates produced by CTFFIND4 (version 4.1.13) and Gctf (version 1.06). 
We denote the vector of estimated defocus parameters produced by CTFFIND4 when using block of size $512 \times 512$ and $1024 \times 1024$ as $\phi_{c}^{(512)}$ and  $\phi_{c}^{(1024)}$, respectively. We further denote the vector of estimated defocus parameters produced by Gctf when using block of size $512 \times 512$ and $1024 \times 1024$ as $\phi_{g}^{(512)}$ and  $\phi_{g}^{(1024)}$, respectively. For each estimation method we select the vector of estimated defocus parameters that leads to highest correlation with the estimated power spectrum, that is
\begin{equation}
\phi_j^* = \arg \underset{\phi_j \in \{\phi_j^{(512)},  \phi_j^{(1024)}  \}}{\max} \left( \frac{1}{s_t} \sum_{m=1}^{s_t} P_{cc}^{m} (\phi_j ) \right),
\end{equation}
where $j \in \{ a, c, g\}$, $s_t$ is the number of micrographs in the $t$th dataset and $P_{cc}^{m}$ is the correlation for the $m$th micrograph in the dataset, computed as in~\eqref{equ:corr_def}. We note that the correlation is computed with the power spectrum estimate suggested in \citep{mindell2003ctffind3}. 
 That is, we compute the Pearson correlation coefficient between the background subtracted power spectrum computed as in~\citep{mindell2003ctffind3} and using blocks of size $512 \times 512$ with $H_{\phi_j^{(512)}}$, and between the background subtracted power spectrum computed using blocks of size $1024 \times 1024$ with $H_{\phi_j^{(1024)}}$. In this manner, we choose the block size that best captures the oscillations of each dataset. 

In order to compare the consistency of these $3$ methods, we present the differences between $\phi_a$, $\phi_c$ and $\phi_g$ in Tables~\ref{tab:mean}-\ref{tab:std}. That is, for each micrograph $m$ we compute 
\begin{equation}
\begin{aligned}
& \epsilon_{j,k}(\dfone) = (\phi_j(1) - \phi_k(1)) / \phi_j(1)\\
& \epsilon_{j,k}(\dftwo) = (\phi_j(2) - \phi_k(2)) / \phi_j(2)\\
& \epsilon_{j,k}(\alpha_f) = (\phi_j(3) - \phi_k(3)) / \phi_j(3)
\end{aligned}
\label{equ:dif_comparison}
\end{equation}
where $j$ and $k$ are two estimation methods (ASPIRE-CTF, CTFFIND4 or Gctf). We report the mean and variance of $\epsilon$.  We note that either $\epsilon_{a,g}$ or  $\epsilon_{a,c}$ are often smaller than $\epsilon_{c,g}$, thus showing the ASPIRE-CTF estimate to be in the consensus of the three estimation vectors.

\begin{table*}
\begin{center}
\scalebox{0.8}{
\begin{tabular}{ | c | c | c | c | c | c | c | c | c | c | c |}
  \hline			
 Dataset & Molecule & $\epsilon_{a,g} (\dfone)$ & $\epsilon_{a,c} (\dfone)$ & $\epsilon_{c,g} (\dfone)$ & $\epsilon_{a,g} (\dftwo)$ &    $\epsilon_{a,c} (\dftwo)$ &   $\epsilon_{c,g} (\dftwo)$ &   $\epsilon_{a,g} (\alpha_f)$ & $\epsilon_{a,c} (\alpha_f)$ &  $\epsilon_{c,g} (\alpha_f)$ \\
 \hline
001 & GroEL & 0.0157 & 0.2370 & 0.2364 & 0.0155 &0.1282 & 0.6081 & 1.6599 &  2.1429 & 1.3848\\
002 & GroEL  & 1.3236 & 1.6500 & 0.1515 & 1.4876 & 0.9746 & 0.2374 & 3.6137 & 5.3979 & 1.9592\\
003 & 60S ribosome & 0.0029 & 0.0049 & 0.0029 & 0.0030 & 0.0032 & 0.0017 &  6.1042 & 6.6618 & 2.9159 \\
004 & 60S ribosome & 0.0051 & 0.0130 & 0.0098 &  0.0057 &  0.0406 & 0.2990 &  4.2653 & 2.3667 & 2.6050\\
005 & apoferritin & 0.0036 &  0.1744 & 0.8809 & 0.0036 & 0.1997 & 1.5230 & 0.6597 & 0.8927 &  0.7413\\
006 & apoferritin & 0.0099 & 0.2094 & 0.6392 & 0.0069 & 0.3409 & 1.5217 & 0.2670 & 1.3014 & 1.6142\\
007 & TMV virus & 0.0224 &  0.2144 & 0.5163 & 0.0265 & 0.2674 & 1.1320 & 0.4254 & 0.9891 & 0.8985 \\
008 & TMV virus & 0.1961 & 0.6043 & 0.3988 & 0.2163 & 0.3016 & 0.5707 & 3.0253 & 1.8514 & 2.8394 \\
\hline
\end{tabular} }
\end{center}
\caption{Comparison between parameters estimated by ASPIRE-CTF, CTFFIND4 and Gctf. We present the mean (over each dataset) of normalized differences between each two CTF estimation methods as detailed in~\eqref{equ:dif_comparison}. The subscripts $a$, $g$ and $c$ indicate ASPIRE-CTF, Gctf and CTFFIND4, respectively.}
\label{tab:mean}
\end{table*}

\begin{table*}
\begin{center}
\scalebox{0.8}{
\begin{tabular}{ | c | c | c | c | c | c | c | c | c | c | c |}
  \hline			
 Dataset & Molecule & $\epsilon_{a,g} (\dfone)$ & $\epsilon_{a,c} (\dfone)$ & $\epsilon_{c,g} (\dfone)$ & $\epsilon_{a,g} (\dftwo)$ &    $\epsilon_{a,c} (\dftwo)$ &   $\epsilon_{c,g} (\dftwo)$ &   $\epsilon_{a,g} (\alpha_f)$ & $\epsilon_{a,c} (\alpha_f)$ &  $\epsilon_{c,g} (\alpha_f)$ \\
 \hline
001 & GroEL  & 0.0144 &  0.4710 & 0.5276 & 0.0267 &  0.2489 & 1.9206  &  3.1585 & 2.0435 & 2.2632\\
002 & GroEL  & 2.9445 & 2.2763 & 0.2655 & 3.3142 & 1.9597 & 0.5196 & 9.2965 & 15.0486 & 2.7796\\
003 & 60S ribosome & 0.0022 &  0.0031 & 0.0036 & 0.0031 &  0.0047 & 0.0020 & 20.8434 & 22.3327 &  15.8681\\
004 & 60S ribosome & 0.0028 & 0.0328 & 0.0387 & 0.0039 & 0.1778 & 1.4504 & 9.4786 & 3.5234 & 3.9324\\
005 & apoferritin & 0.0037 & 0.3265 & 1.8456 & 0.0029 & 0.3591 &  3.1834 & 0.9381 & 0.7127 & 0.6918\\
006 & apoferritin & 0.0126 & 0.2902 & 1.9539 & 0.0064 & 0.3330 &  2.7270 & 0.3485 & 1.1711 & 2.7480 \\
007 & TMV virus & 0.0192 & 0.2698 & 0.8581 & 0.0172  & 0.3335 & 2.0805 & 0.6437 & 0.9296 & 0.9966 \\
008 & TMV virus & 0.8337 & 0.7936 & 0.8543 & 0.8947 & 0.4088 &  0.9907 & 4.7943 & 2.6911 &  6.1583 \\
\hline
\end{tabular}
}
\end{center}
\caption{Comparison between parameters estimated by ASPIRE-CTF, CTFFIND4 and Gctf. We present the standard deviation (over each dataset) of normalized differences between each two CTF estimation methods as detailed in~\eqref{equ:dif_comparison}. The subscripts $a$, $g$ and $c$ indicate ASPIRE-CTF, Gctf and CTFFIND4, respectively.
}
\label{tab:std}
\end{table*}

\begin{centering}
\begin{figure}[t]
\centering
{\label{subfig:psd}\includegraphics[width=0.6\linewidth]{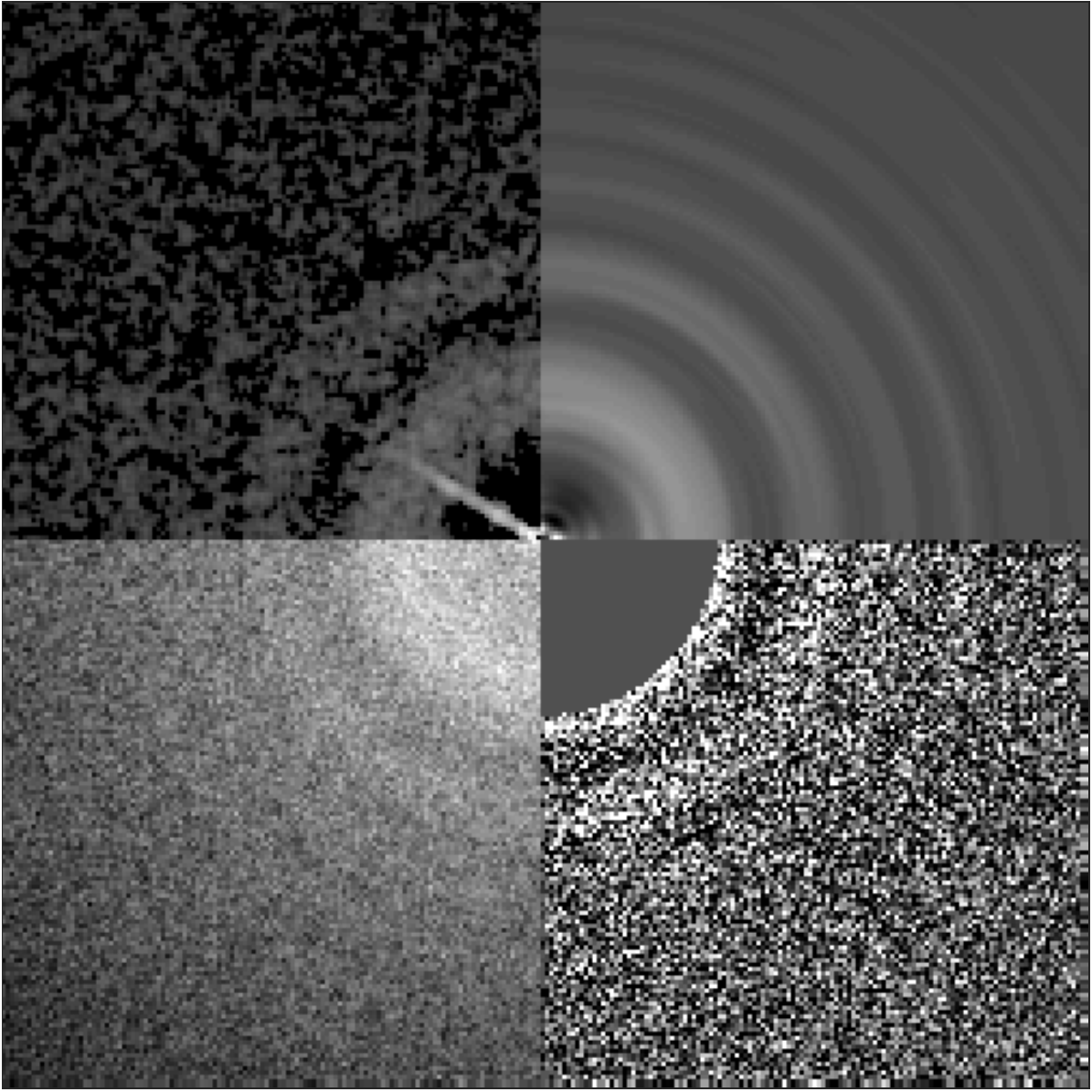}}
\caption{%
Visual comparison between the power spectra computed by ASPIRE-CTF (top row), Gctf (bottom left) and CTFFIND4 (bottom right) on a sample micrograph of dataset 008. On the top row we present $\hpsdlp_\z$ (right) and $\hpsd_\z$ (left).}
\label{fig:vis_compare}
\end{figure}
\end{centering}

Fig.~\ref{fig:vis_compare} contains a visual comparison between the power spectrum computed by our suggested framework and the power spectra computed by Gctf and CTFFIND4. We present the comparison over a micrograph from the eighth set of the CTF challenge as this set is known to be difficult. We note that the oscillations of the ASPIRE-CTF power spectrum are highly noticeable. In comparison, the variability of the power spectra computed by Gctf and CTFFIND4 make visual detection of oscillations challenging.  

\subsection{Runtime}

We compute runtime
of ASPIRE-CTF and  CTFFIND4 over dataset $001$ of the CTF challenge. For both methods, we partition the micrograph into blocks of size  $512 \times 512$. When running ASPIRE-CTF we employ $L=4$ data tapers. Furthermore, we use the exhaustive search option for CTFFIND4, and perform an exhaustive 1D search in ASPIRE-CTF. While the CTF estimation results are comparable, there is a significant speedup when using ASPIRE-CTF. Runtime for ASPIRE-CTF is 22.5 seconds on average per micrograph, while runtime for CTFFIND4 is 541 seconds.

These experiments are run on a $2.6$ GHz Intel Core i7 CPU with four cores and $16$ GB of memory. We do not compare to the runtime of Gctf as it must be run on a GPU.

\subsection{Consistency in low SNR}

To test consistency of results with changing SNR, we turn to the EMPIAR-10249 dataset \citet{herzik2019dataset}.
This dataset consists of movies with $44$ frames per movie.
Usually, all these frames, except for a few frames at the beginning and a few at the end, are motion-corrected and summed to create a micrograph.
This is due to the fact that a micrograph created from as many motion-corrected frames as possible will have the best SNR.

We disregard the first frame and use MotionCor2 \citep{Zheng2017motion} to create $9$ motion-corrected micrographs.
These consist of summing $5$, $8$, $13$, $18$, $23$, $28$, $33$, $38$, and $43$ motion-corrected frames, respectively.
This gives us a sequence of micrographs with increasing SNR.

\begin{figure}[t]
\centering
{\label{subfig:psd}\includegraphics[width=0.65\linewidth]{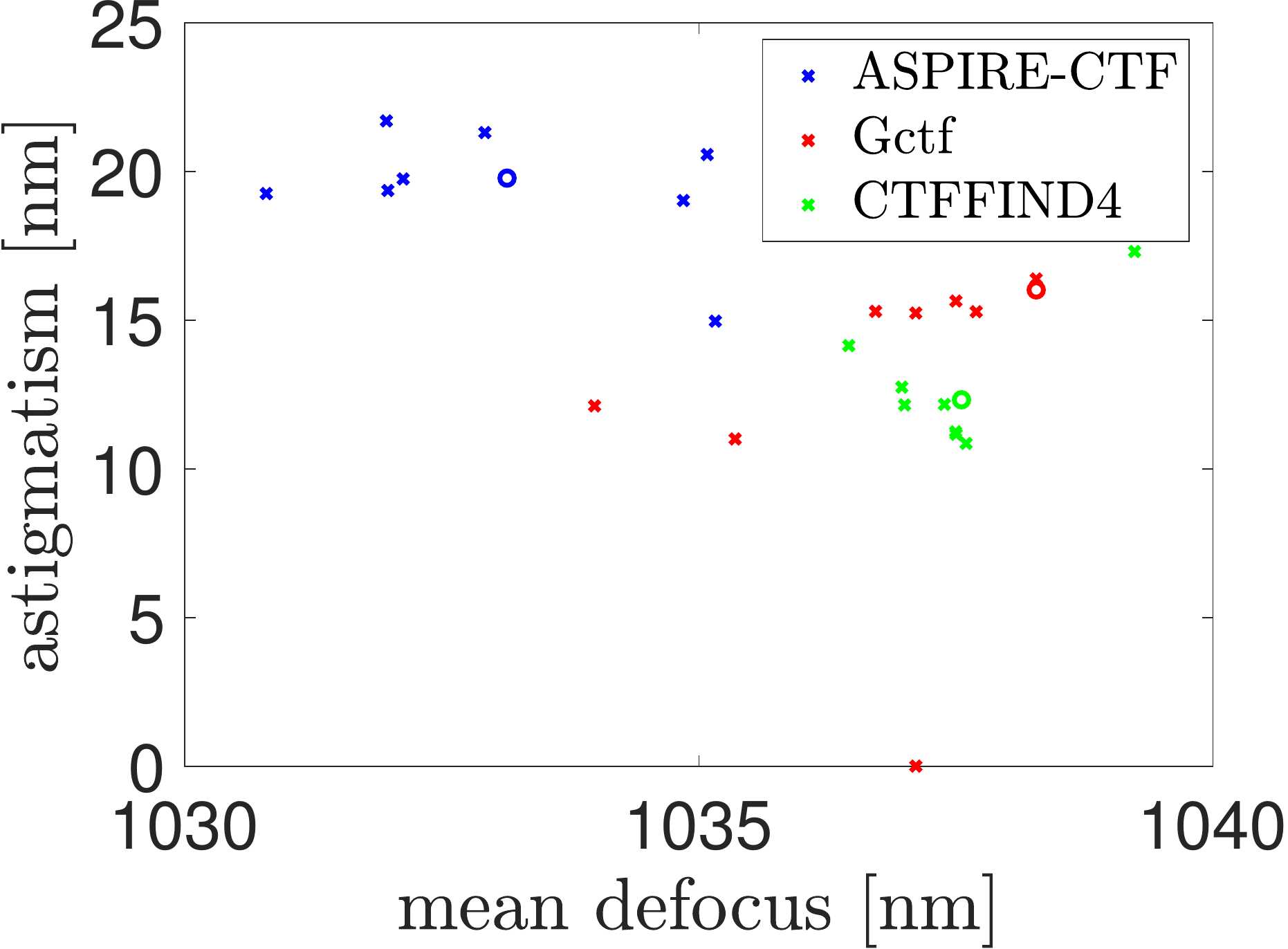}}
\caption{Estimated astigmatism vs. defocus of the CTF parameters. The circular markers present the defocus and astigmatism estimated from a micrograph with $43$ summed frames.}
\label{fig:astigmatism_vs_defocus}
\end{figure}

We estimated the CTF parameters independently from each micrograph in the manner detailed in Section~\ref{subsec:exp1}. 
Fig. \ref{fig:astigmatism_vs_defocus} shows the astigmatism $|\dfone-\dftwo|$ vs.\ mean defocus $(\dfone + \dftwo)/2$ of the CTF estimation for each method and over each micrograph. We see that while the average defocus values remain similar for all three methods, Gctf incurs a larger error in the astigmatism when 23 frames are used. On the other hand, our method and CTFFIND4 achieve consistent estimates regardless of the amount of frames averaged.

\section{Conclusion}

In this paper we have presented a novel approach for power spectrum estimation of cryo-EM experimental data. Our approach uses the multitaper estimator, which often leads to reduced 
mean square error over Bartlett's and Welch's methods.
Additionally, we presented a method for error reduction that is driven directly by the mathematical model of the contrast transfer function. We did this by projecting the power spectrum estimate onto a steerable basis and discarding any basis function where the CTF must be negligible. We showed that the combination of these two contributions leads to greatly reduced variability in our estimator. 

We presented experimental results on twelve datasets, and showed that our method is well
suited to both motion-corrected micrographs and raw movies data.

\section*{Acknowledgments}
This work was partially supported by the Simons Foundation Math+X Investigator Award and the Moore Foundation Data-Driven Discovery Investigator Award. 
The authors %would like to 
thank B. Landa and I. Sason for help optimizing the PSWF code.
The authors are also indebted to B. Landa, Y. Shkolnisky and A. Rohou for helpful comments and discussions.
The Flatiron Institute is a division of the Simons Foundation.

\section*{Appendix A}

Zeroth-order discrete prolate spheroidal sequence (DPSS) \citep{slepian1978prolates} were proposed as data tapers in (\citealp{babadi2014multitaper}; Section~\ref{subsec:bias}). Here we describe their generation. 

The zeroth-order discrete prolate spheroidal sequence is a sequence of $d$ 1-D vectors, determined as the $d$ leading eigenvectors of the matrix $\mathbf{\mathcal{L}} \in \mathbb{R}^{K \times K}$, where 
\begin{equation}
\mathcal{L}[{k,m}] = \frac{\sin (\pi R (k-m)))}{\pi (k-m)},
\end{equation}
and $R = \frac{2d}{N}$. We denote the resulting data tapers by $\mathbf{t}_1,\dots,\mathbf{t}_N$. 

As the blocks $\mathbf{y}_b$ are 2D, that is $\mathbf{y}_b \in \mathbb{R}^{K \times K}$, the data tapers we use are defined as
$$\mathbf{w}_{pq} = \mathbf{t}_p^T \mathbf{t}_q, \quad 0 \le p,q < d.$$
$$w_{dq+p} [k_1, k_2] = t_p[k_1] t_q[k_2].$$

Lastly, we note that $d$ is selected such that
$$ (d-1)^2 < L \le d^2.$$

%\FloatBarrier
\section*{References}
\bibliography{ctf}
%\end{multicols}
\end{document}